\documentclass[11pt]{article}%
\usepackage{amssymb}
\usepackage{amsfonts}
\usepackage{amsmath}
\usepackage{authblk}
\usepackage{url}
\usepackage{hyperref}
\usepackage{wrapfig}
\usepackage{xcolor}
\usepackage{soul}
\usepackage{graphicx}%
\usepackage{mwe}
\usepackage{subcaption}
\usepackage{lineno}
\usepackage{algpseudocode,algorithm,algorithmicx}
\providecommand{\U}[1]{\protect\rule{.1in}{.1in}}
\newtheorem{theorem}{Theorem}

\newtheorem{property}{Property}

\begin{document}

\title{Linear Clustering Process on Networks}

\author[1]{\small Ivan Joki\'{c}\thanks{\emph{I.Jokic@tudelft.nl}}}
\author[1]{\small Piet Van Mieghem}

\affil[1]{\footnotesize Faculty of Electrical Engineering, Mathematics, and Computer Science,
Delft University of Technology, P.O. Box 5031, 2600 GA Delft, The Netherlands}

\maketitle

\begin{abstract}
We propose a linear clustering process on a network consisting of two opposite forces: attraction and repulsion between adjacent nodes. Each node is mapped to a position on a one-dimensional line. The attraction and repulsion forces move the nodal position on the line, depending on how similar or different the neighbourhoods of two adjacent nodes are. 
Based on each node position, the number of clusters in a network, together with each node's cluster membership, is estimated.
The performance of the proposed linear clustering process is benchmarked on synthetic networks against widely accepted clustering algorithms such as modularity, the Louvain method and the non-back tracking matrix. The proposed linear clustering process outperforms the most popular modularity-based methods, such as the Louvain method, while possessing a comparable computational complexity.
\end{abstract}

\section{Introduction}\label{Sec:Introduction}

Networks \cite{barabasi2013network,newman2018networks} abound and increasingly shape our world, ranging from infrastructural networks (transportation, telecommunication, power-grids, water, etc.) over social networks to brain and biological networks. In general, a network consists of a graph or underlying topology and a dynamic process that takes place on the network. Some examples of processes on a network are percolation \cite{karrer2014percolation} and epidemic spreading \cite{prasse2019viral,pastor2015epidemic}, that possess a phase transition \cite{sanchez2002nonequilibrium,stanley1971phase}. While most real-world processes on networks are non-linear, linearisation allows for hierarchical structuring of processes on the network \cite{jokic2020linear}.

The identification of communities and the corresponding hierarchical structure in real-world networks has been an active research topic for decades \cite{fortunato2010community}, although a single, precise definition of a community does not seem to exist \cite{budel2020detecting,newman2012communities}. In network science, a community is defined as a set of nodes that share links dominantly between themselves, while a minority of links is shared with other nodes in the network. 
Newman proposed in \cite{newman2006modularity} a spectral clustering algorithm that reveals hierarchical structure of a network, by optimising modularity, a commonly used quality function of a graph partition. A heuristic, modularity-based two-step clustering algorithm, proposed by Blondel \textit{et al.} in \cite{blondel2008fast}, has proved to be computationally efficient and performed among the best in the comparative study conducted in \cite{lancichinetti2009community}. Recently, Peixoto proposed in \cite{peixoto2014hierarchical} a nested generative model, able to identify nested partitions at different resolutions, which thus overcomes an existing drawback of a majority of clustering algorithms, identifying small, but well-distinguished communities in a large network. Dannon \textit{et al.} concluded in their comparative study \cite{danon2005comparing} that those clustering algorithms performing the best tend to be less computationally efficient.
We refer to \cite{fortunato2010community} for a detailed review on existing clustering algorithms.

Our new idea here is the proposal of a linear clustering process (LCP) on a graph, where nodes move in one-dimensional space and tend to concentrate in groups that lead to network communities. Linear means "proportional to the graph", which is needed, because the aim is to cluster the graph and the process should only help and not distract from our main aim of clustering. A non-linear process depends intricately on the underlying graph that we want to cluster and may result in worse clustering! Our LCP leads to a new and non-trivial graph matrix $W$ in (\ref{def_martrix_W}) in Theorem \ref{clustering_law_additive_matrix_form}, whose spectral decomposition is at least as good as the best clustering result, based on the non-back tracking matrix \cite{krzakala2013spectral}. Moreover, the  new graph matrix $W$ has a more "natural" relation to clustering than the non-back tracking matrix, that was not designed for clustering initially. Finally, our resulting LCP clustering algorithm seems surprisingly effective and can compete computationally with any other clustering algorithm, while achieving generally a better result!

In Section \ref{Sec:Network_Clustering}, we introduce notations for graph partitioning and briefly review basic theory on clustering such as modularity and the stochastic block model. We introduce the linear clustering process (LCP) on network in Section \ref{Sec_dynamic_clustering}, while the resulting community detection algorithm is described in Section \ref{Sec:Position_Vector_Analysis} and Section \ref{Sec_Iterative_Computation}. We compare the performance of our LCP algorithm with that of the non-back tracking matrix, Newman's and the Louvain algorithm and provide results in Section \ref{Sec_Results}, after which we conclude.

\section{Network or graph clustering}\label{Sec:Network_Clustering} 

A graph $G\left(  \mathcal{N},\mathcal{L}\right)  $ consists of
a set $\mathcal{N}$ of $N=\left\vert \mathcal{N}\right\vert $ nodes and a set $\mathcal{L}$ of $L=\left\vert \mathcal{L}\right\vert $ links and is defined by the $N\times N$ adjacency matrix $A$, where $a_{ij}=1$ if node $i$ and node $j$ are connected by a link, otherwise $a_{ij}=0$. The $N\times1$ degree vector $d$ obeys $d=A\cdot u$, where the $N\times1$ all-one vector $u$ is composed of ones. The corresponding $N\times N$ degree diagonal matrix is denoted by $\Delta=\text{diag}\left(  d\right)  $. 

The set of neighbours of node $i$ is denoted by $\mathcal{N}_{i}=\{k\mid
a_{ik}=1,k\in\mathcal{N}\}$ and the degree of node $i$ equals the cardinality
of that set, $d_{i}=\left\vert \mathcal{N}_{i}\right\vert $. The set of common
neighbours of node $i$ and node $j$ is $\mathcal{N}_{i}\cap\mathcal{N}_{j}$,
while the set of neighbours of node $i$ that do not belong to node $j$ is
$\mathcal{N}_{i}\setminus\mathcal{N}_{j}$. The degree of a node $i$ also
equals the sum of the number of common and different neighbours between nodes
$i$ and $j$
\begin{equation}
	d_{i}=\left\vert \mathcal{N}_{i}\setminus\mathcal{N}_{j}\right\vert
	+\left\vert \mathcal{N}_{i}\cap\mathcal{N}_{j}\right\vert
	\label{eq_degree_in_set_notation}%
\end{equation}
The number of common neighbours between nodes $i$ and $j$ equals the $ij$-th
element of the squared adjacency matrix
\begin{equation}
	\left\vert \mathcal{N}_{i}\cap\mathcal{N}_{j}\right\vert =\left(
	A^{2}\right)  _{ij} \label{common_neig_set_not}%
\end{equation}
because $(A^{k})_{ij}$ represents the number of walks with $k$ hops between
node $i$ and node $j$ (see \cite[p. 32]{van2010graph}). From
(\ref{eq_degree_in_set_notation}), (\ref{common_neig_set_not}) and
$d_{i}=(Au)_{i}=(A^{2})_{ii}$, we have
\[
\left\vert \mathcal{N}_{i}\setminus\mathcal{N}_{j}\right\vert =\left(
A^{2}\right)  _{ii}-\left(  A^{2}\right)  _{ij}%
\]
and
\[
\left\vert \mathcal{N}_{i}\setminus\mathcal{N}_{j}\right\vert +\left\vert
\mathcal{N}_{j}\setminus\mathcal{N}_{i}\right\vert =\left(  A^{2}\right)
_{ii}+\left(  A^{2}\right)  _{jj}-2\left(  A^{2}\right)  _{ij}%
\]
The latter expression is analogous to the effective resistance $\omega_{ij}$
between node $i$ and node $j$,
\[
\omega_{ij}=Q_{ii}^{\dag}+Q_{jj}^{\dag}-2Q_{ij}^{\dag}
\]
in terms of the pseudoinverse $Q_{ii}^{\dag}$ of the Laplacian matrix $Q=$
$\Delta-A$ (see e.g. \cite{van2017pseudoinverse}).

Network clustering aims to organize nodes in the graph $G$ in such a way that
the majority of links in the graph connect nodes of the same cluster, while a
minority of links is shared by nodes from different clusters. This general
description is imprecise and part of the complication, which causes that
community detection in a network is a challenging problem, that has been
actively researched over the last few decades \cite{fortunato2010community}.

Before introducing our linear clustering process (LCP) in Section \ref{Sec_dynamic_clustering}, we briefly present basic graph partitioning concepts, while the overview of the more popular clustering methods is deferred to Appendix \ref{App_Clustering_Algorithms}.

\subsection{Network modularity}\label{Sec_2_1} 

Newman and Girvan \cite{newman2004finding} proposed the modularity as a concept for a network partitioning,
\begin{equation}\label{Eq_modularity}
	m = \frac{1}{2L}\cdot \sum\limits_{i=1}^{N} \sum\limits_{j=1}^{N} \left(a_{ij}-\frac{d_{i}\cdot d_{j}}{2L}\right) \cdot  \mathbf{1}_{\{i\, \text{and}\, j \,\in
		\text{ same cluster}\}},
\end{equation}
where $\mathbf{1}_{x}$ is the indicator function that equals 1 if statement $x$ is true, otherwise $\mathbf{1}_{x}=0$. The modularity $m$ compares the number of links between nodes from the same community with the expected number of intra-community links in a network with randomly connected nodes. When the modularity $m$ close to $0$, the estimated partition is as good as a random partition would be. On the contrary, a modularity $m$ close to $1$ indicates that the network can be clearly partitioned into clusters. Defining the $N\times N$ modularity matrix $C$,
\begin{equation}\label{Modularity_matrix_C}
	C_{ij}=	\begin{cases}
		1 & \text{if nodes $i$ and $j$ belong to the same cluster}\\
		0 & \text{otherwise},
	\end{cases}
\end{equation}
allows us to rewrite the modularity (\ref{Eq_modularity}) as a quadratic form,
\begin{equation}\label{Eq_modularity_m_using_C}
	m =  \frac{1}{2L}\cdot u^{T}\cdot \left(A\circ C - \frac{1}{2L}\cdot \left(d\cdot d^{T}\right)\circ C\right)\cdot  u,
\end{equation}
where $\circ$ denotes the Hadamard product \cite{horn2012matrix}. The number of clusters in a network is denoted by $c$, while the $c\times 1$ vector $n$ defines the size of each cluster:
\begin{equation}\label{Eq:Cluster_Size}
	n = \begin{bmatrix}
		n_{1} & n_{2} & \dots & n_{c}
	\end{bmatrix},
\end{equation}
where the number of nodes in cluster $i$ is denoted as $n_{i}$.

\subsection{Stochastic block model}\label{SubSec:SBM}

The performance of the clustering methods in this paper are benchmarked on random graphs, generated by the Stochastic Block Model (SBM), proposed by Holland \cite{holland1983stochastic}. The SBM model generates a random graph with community structure, where a link between two nodes exists with different probability, depending on whether the nodes belong to the same cluster or not. 

In this paper, we focus on the symmetric stochastic block model (SSBM), where only two different link probabilities are defined. Two nodes are connected via a link with probability $p_{in}$ if they belong to the same cluster, otherwise the direct link exists with probability $p_{out}$. Communities emerge when the link density within clusters is larger than the inter-community link probability $p_{in}>p_{out}$. Furthermore, we restrict clusters to be of the same size: \[n_{i} = \frac{N}{c}, i\in \{1,\,2,\,\dots \,,c\},\]
which causes the expected degree to be the same for each node:
\begin{equation}\label{Eq:Expected_Degree}
	E[D] = \frac{b_{in}+\left(c-1\right)\cdot b_{out}}{c},
\end{equation}
irrespective of its cluster membership. We further consider a sparse and assortative SSBM. The SBMM is called sparse and assortative when the link probabilities $p_{in}=\frac{b_{in}}{N}$ and $p_{out}=\frac{b_{out}}{N}$ are defined upon positive constants $b_{in}>b_{out}$ that stay constant when $N \rightarrow \infty$. Decelle \textit{et al.} \cite{decelle2011inference,decelle2011asymptotic} found that when the difference $b_{in}-b_{out}$ is above the detectability threshold
\begin{equation}\label{Eq:SBM_treshold}
	b_{in} - b_{out} > c\cdot \sqrt{E[D]},
\end{equation}
it is theoretically possible to recover cluster membership of the nodes, otherwise, the community structure of a network is not distinguishable from randomness. The threshold (\ref{Eq:SBM_treshold}) marks a phase transition between the undetectable and the theoretically detectable regime of the SSBM.

\section{Linear clustering process (LCP) on a graph}
\label{Sec_dynamic_clustering}

\subsection{Concept of the clustering process}

Each node $i$ in the graph $G$ is assigned a position $x_{i}[k]$ on a line
(i.e. in one-dimensional space) at discrete time $k$. We define the $N\times1$
position vector $x[k]$ at discrete time $k$, where the $i$-th vector component
consists of the position $x_{i}[k]$ of node $i$ at time $k$. We initialize the
$N\times1$ position vector $x[0]$ by placing nodes equidistantly on the line
and assign integer values from $1$ to $N$ to the nodes, thus, $x[0]=%
\begin{bmatrix}
	1 & 2 & \dots & N
\end{bmatrix}
^{T}$. At last, we restrict the position $x_{i}[k]$ to positive real values.

We propose a dynamic process that determines the position of nodes over time.
The position difference between nodes of the same cluster is relatively small.
On the contrary, nodes from different clusters are relatively far away, i.e.
their position difference is relatively high. Based on the position vector
$x[k]$, we will distinguish clusters, also called communities, in the graph
$G$.

The proposed clustering process consists of two opposite and simultaneous
forces that change the position of nodes at discrete time $k$:

\begin{description}
	\item[Attraction] Adjacent nodes sharing many neighbours are
	mutually attracted with a force proportional to the number of common
	neighbours. In particular, the attractive force between node $i$ and its
	neighboring node $j$ is proportional to $\alpha\cdot\left(  \left\vert
	\mathcal{N}_{j}\cap\mathcal{N}_{i}\right\vert +1\right)  $, where $\alpha$ is
	the attraction strength and $\left(  \left\vert \mathcal{N}_{j}\cap
	\mathcal{N}_{i}\right\vert +1\right)  $ equals the number of common neighbors
	plus the direct link, i.e. $a_{ij}=1$. 
	
	\item[Repulsion] Adjacent nodes sharing a few neighbours are
	repulsed with a force proportional to the number of different neighbours. The
	repulsive force between node $i$ and its neighboring node $j$ is proportional
	to $\delta\cdot\left(  \left\vert \mathcal{N}_{j}\setminus\mathcal{N}%
	_{i}\right\vert -1\right)  $, where $\delta$ is the repulsive strength and
	$\left(  \left\vert \mathcal{N}_{j}\setminus\mathcal{N}_{i}\right\vert
	-1\right)  $ equals the set of neighbours of node $j$ that do not belong to
	node $i$ minus the direct link (that is included in $\left\vert \mathcal{N}%
	_{j}\setminus\mathcal{N}_{i}\right\vert )$. Since the force should be
	symmetric and the same if $i$ and $j$ are interchanged, we end up with a
	resultant repulsive force proportional to $\frac{1}{2}\cdot\delta\cdot\left(
	\left\vert \mathcal{N}_{j}\setminus\mathcal{N}_{i}\right\vert +\left\vert
	\mathcal{N}_{i}\setminus\mathcal{N}_{j}\right\vert -2\right)$.  
\end{description}

\subsection{LCP in discrete time}

Since computers operate with integers and truncated real numbers, we concentrate
on discrete-time modeling. The continuous-time description is derived in Appendix \ref{Sec_Continuous_time}. We denote the continuous-time
variables by $y\left(  t\right)  $ and the continuous time by $t$, while the
discrete-time counterpart is denoted by $y\left[  k\right]  $, where the
integer $k$ denotes the discrete time or $k$-th timeslot. The transition from
the continuous-time derivative to the discrete-time difference is
\begin{eqnarray*}
	\frac{dx_{i}(t)}{dt} =  \lim_{\Delta t\rightarrow 0}\frac{x_{i}(t+\Delta t)-x_{i}(t)}{\Delta t}
	\rightarrow &
	\left.\frac{x_{i}(t+\Delta t)-x_{i}(t)}{\Delta t}\right\vert_{\Delta t=1} \overset{\text{def}}{=}  x_{i}\left[k+1\right]  -x_{i}\left[k\right]
\end{eqnarray*}
Corresponding to the continuous-time law in Appendix \ref{law_continuous_time} and
choosing the time step $\Delta t=1$, the governing equation of position
$x_{i}[k]$ of node $i$ at discrete time $k$ is
\begin{equation}\label{clustering_law_discrete_time}%
x_{i}[k+1]= x_{i}[k]+\sum\limits_{j\in\mathcal{N}_{i}} \bigg(\frac{\alpha\cdot\left(\left\vert \mathcal{N}_{j}\cap\mathcal{N}_{i}\right\vert +1\right)}{d_{j}d_{i}}- \frac{\frac{1}{2}\cdot\delta\cdot\left(  \left\vert \mathcal{N}_{j}\setminus \mathcal{N}_{i}\right\vert +\left\vert \mathcal{N}_{i}\setminus\mathcal{N}_{j}\right\vert -2\right)}{d_{j}d_{i}} \bigg)\cdot\Big(x_{j}[k]-x_{i}[k]\Big)
\end{equation}
where $\alpha$ and $\delta$ are, in the discrete-time setting, the strength
(in dimensionless units) for attraction and repulsion, respectively. The
maximum position difference at the initial state is $x_{N}[0]-x_{1}[0]=N-1$.

Node $j$ attracts an adjacent node $i$ with force proportional to their
position difference ($x_{j}[k]-x_{i}[k]$). The intensity of the attractive
force decreases as nodes $i$ and $j$ are closer on a line. The attraction is
also proportional to the number common neighbours $\left\vert \mathcal{N}%
_{j}\cap\mathcal{N}_{i}\right\vert $ of node $i$ and node $j$ plus the direct
link, as nodes tend to share most links with other nodes from the same
cluster. On the contrary, node $j$ repulses node $i$ with a rate proportional
to their position difference ($x_{j}[k]-x_{i}[k]$) and the average of the
number of node $j$ neighbours $\left\vert \mathcal{N}_{j}\setminus
\mathcal{N}_{i}\right\vert $ that are not connected to the node $i$ and,
similarly, the number of node $i$ neighbors, $\left\vert \mathcal{N}%
_{j}\setminus\mathcal{N}_{i}\right\vert $ that are not connected to the node
$j$. The repulsive and attractive force are, as mentioned above, symmetric in
strength, but opposite, if $i$ is interchanged by $j$.

\begin{figure*}[!]
	\begin{center}
		\includegraphics[angle=0, scale=0.7]{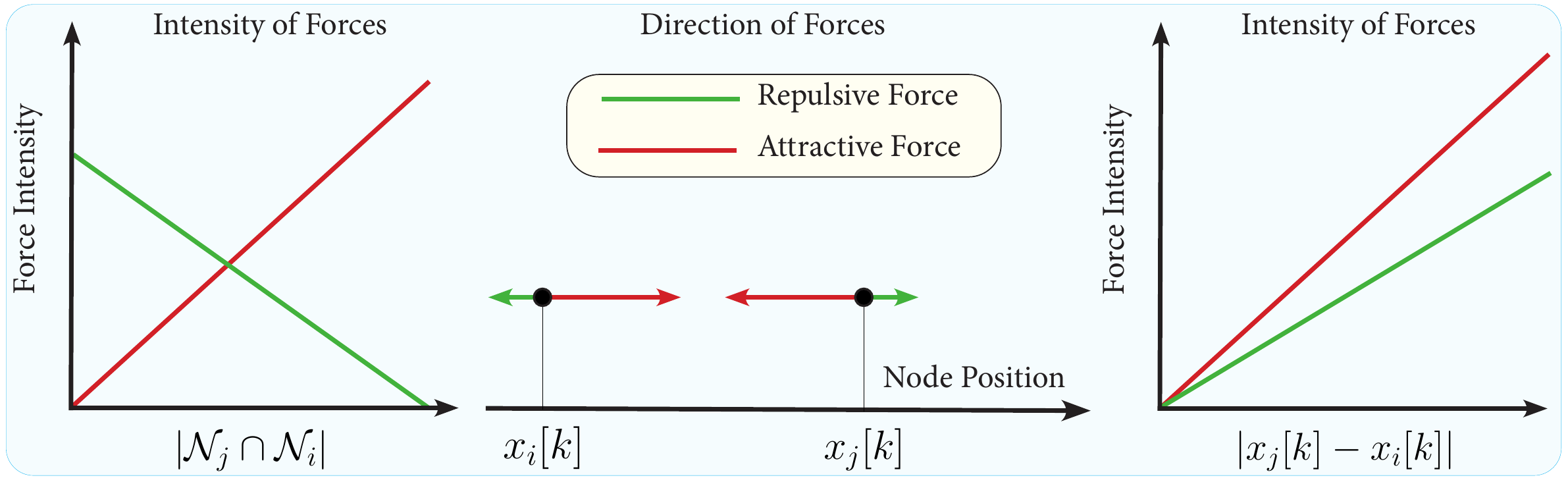}
		\caption{Dependence of the attractive and repulsive force on the number of common neighbours of adjacent nodes $i$ and $j$ (left-figure). Directions of the attraction and repulsion forces between the adjacent nodes (middle-figure). Dependence of the attractive and repulsive force on the absolute position distance between adjacent nodes $i$ and $j$ (right-figure).}
		\label{Fig_Forces_1}
	\end{center}
\end{figure*}

The directions of both attractive and repulsive forces between two adjacent
nodes $i$ and $j$ as well the dependence of both forces on the number of
common neighbours $\left|  \mathcal{N}_{j}\cap\mathcal{N}_{i}\right|  $ and
the absolute position distance $\left|  x_{j}[k]-x_{i}[k] \right|  $ are
illustrated in Fig \ref{Fig_Forces_1}.

In the continuous-time setting, as provided in Appendix \ref{law_continuous_time}, we eliminate one parameter by scaling the time $t^{\ast}=\delta t$. Because the time step
$\Delta t=1$ is fixed and cannot be scaled, the discrete-time model
consists of two parameters $\alpha\geq0$ and $\delta\geq0$.

So far, we have presented an additive law, derived in the common Newtonian
approach. The corresponding multiplicative law in discrete time is%

\begin{equation}\label{clustering_law_multiplicative}
	\resizebox{0.9 \textwidth}{!}{$
			x_{i}[k+1] = x_{i}[k]\cdot \Bigg( 1+ \sum\limits_{j\in\mathcal{N}_{i}} \bigg(\frac{\alpha\cdot\left(  \left| \mathcal{N}_{j} \cap\mathcal{N}_{i} \right|  + 1\right)}{d_{i}\cdot d_{j}}  - \frac{\frac{1}{2} \cdot\delta\cdot\left(  \left|  \mathcal{N}_{j} \setminus\mathcal{N}_{i} \right|  + \left|  \mathcal{N}_{i} \setminus\mathcal{N}_{j} \right|  -2 \right)}{d_{i}\cdot d_{j}} \bigg)\cdot \Big(x_{j}[k] - x_{i}[k]\Big) \Bigg)$}
\end{equation}

Although the physical intuition is similar, the multiplicative process in
(\ref{clustering_law_multiplicative}) behaves different in discrete time than
the additive law in (\ref{clustering_law_discrete_time}). Since also the
analysis is more complicated, we omit a further study of the multiplicative law.

We present the analogon of (\ref{clustering_law_discrete_time}) in a matrix form:

\begin{theorem}
	\label{clustering_law_additive_matrix_form} The discrete time process
	(\ref{clustering_law_discrete_time}) satisfies the linear matrix difference
	equation
	\begin{equation}
		x[k+1]=\left(  I+W-\text{diag}\left(  W\cdot u\right)  \right)  \cdot
		x[k],\label{clustering_law_matrix_equation}%
	\end{equation}
	where the $N\times1$ vector $u$ is composed of ones, the $N\times N$ identity matrix is denoted by $I$, while the $N\times N$ topology-based matrix $W$ is defined as
	\begin{equation}\label{def_martrix_W}
			W = \left(  \alpha+\delta\right)  \Delta^{-1}\cdot\left(  A\circ A^{2}+A\right) \cdot \Delta^{-1}-  \frac{1}{2}\cdot\delta\left(  \Delta^{-1}\cdot A+A\cdot \Delta^{-1}\right)
	\end{equation}
	where $\circ$ denotes the Hadamard product. In particular,
	\begin{equation} \label{def_matrix_W_element}
			w_{ij}=a_{ij}\frac{\alpha \left(  \left\vert \mathcal{N}_{j} \cap\mathcal{N}_{i}\right\vert +1\right) - \delta \left(  \frac{\left\vert \mathcal{N}_{j}\setminus\mathcal{N}_{i}\right\vert +\left\vert \mathcal{N}_{i}\setminus\mathcal{N}_{j}\right\vert }{2}-1\right) }{d_{i}d_{j}}
	\end{equation}
	The explicit solution of the difference equation
	(\ref{clustering_law_matrix_equation}) is
	\begin{equation}
		x[k]=\left(  I+W-\text{diag}\left(  W\cdot u\right)  \right)  ^{k}%
		x[0]\label{solution_positionvector}%
	\end{equation}
	where the $k$-th component of the initial position vector is $(x[0])_{k}=k$.
	
\end{theorem}

\textit{Proof:} Appendix \ref{App_A}.

Theorem \ref{clustering_law_additive_matrix_form} determines the position of the nodal vector $x[k]$ at time $k$ and shows convergence towards a state, where the sum of attractive and repulsive forces (i.e. the resulting force) acting on a node are in balance. Nodes with similar neighbourhoods are grouped on the line, i.e. in the one-dimensional space, while nodes with a relatively small number of common neighbours are relatively far away. A possible variant
of the proposed linear clustering process may map the nodal position into a
higher dimensional space, like a circular disk or square in two dimensions,
and even with a non-Euclidean distance metric.

\subsection{Time-dependence of the linear clustering process}

\label{model_time_dependence_subsection}

The $N \times N$ matrix $I+W - \text{diag}\left(  W \cdot u \right)  $ in the governing equation (\ref{clustering_law_matrix_equation}) has interesting properties. As shown in this section, the related matrix $W - \text{diag}\left(  W \cdot u \right) $ belongs to the class of $M$-matrices, whose eigenvalues have a non-negative real part. The (weighted) Laplacian is another element of the $M$-matrix class.

\begin{property}
	\label{prop_M_non_negative}
	The matrix $I+W - \text{diag}\left(  W \cdot u \right)  $ is a
	non-negative matrix.
\end{property}
\textit{Proof:} The governing equation (\ref{clustering_law_matrix_equation})
\[
x[k+1] = \left(  I + W - \text{diag}\left(  W \cdot u \right)  \right)  \cdot
x[k]
\]
holds for any non-negative vector $x[k]$. Let $x[0] = e_{m}$, the basic vector
with components $\left(  e_{m}\right)  _{i} = \delta_{mi}$ and $\delta_{mi}$
is the Kronecker delta, then we find that the $m$-th column
\[
x[1] = \left(  I+W - \text{diag}\left(  W \cdot u \right)  \right)  _{col(m)}%
\]
must be a non-negative vector. Since we can choose $m$ arbitrary, we have
established that $I+W - \text{diag}\left(  W \cdot u \right)  $ is a
non-negative matrix. $\hfill\square$ 

\begin{property}
	\label{prop_M_eigenvalues}
	The principal eigenvector of the matrix $I+W - \text{diag}\left(  W \cdot u \right)  $ is the all-one vector $u$ belonging to eigenvalue 1. All other eigenvalues of matrix $I+W - \text{diag}\left(  W \cdot u \right)  $ are real and, in absolute value, smaller than 1.
\end{property}
\textit{Proof:} Appendix \ref{sec_prop_M_eigenvalues}.

The linear discrete-time system in
(\ref{clustering_law_matrix_equation}) converges to a steady-state, provided
that $\lim_{k\rightarrow\infty}||x[k+1]|| = \lim_{k\rightarrow\infty
}||x[k]||=||x_{s}||$, which is only possible if the matrix $\left(  I + W -
\text{diag}\left(  W \cdot u \right)  \right)  $ has all eigenvalues in
absolute value smaller than 1 and the largest eigenvalue is precisely equal to
1. Property \ref{prop_M_eigenvalues} confirms convergence and indicates that the steady-state vector $x_{s} = u$ in which the position of each node is the same. However, the steady state solution
$x_{s}=u$ is a trivial solution, as observed from the governing equation in
(\ref{clustering_law_discrete_time}), because the sum vanishes and the
definition of the steady state tells that $x[k+1]=x[k]$, which is obeyed by
any discrete-time independent vector. In other words, the matrix equation
(\ref{clustering_law_matrix_equation}) can be written as
\[
x[k+1]-x[k]=\left(  W-\text{diag}\left(  W\cdot u\right)  \right)
\cdot\left(  x[k]-u\right)
\]
which illustrates that, if $x[k]$ obeys the solution, then $r[k]=x[k]+s\cdot
u$ for any complex number $s$ is a solution, implying that a shift in the
coordinate system of the positions does not alter the physics.

Let us denote the eigenvector $y_{k}$ belonging to the $k$-th
eigenvalue $\beta_{k}$ of the matrix $W-\text{diag}\left(  W\cdot u\right)  $,
where $\beta_{1}\geq\beta_{2}\geq\cdots\geq\beta_{N}$, then the eigenvalue
decomposition of the real, symmetric matrix is
\[
W-\text{diag}\left(  W\cdot u\right)  =Y\text{diag}(\beta)Y^{T}%
\]
where the eigenvalue vector $\beta=(\beta_{1},\beta_{2},\cdots,\beta_{N})$ and
$Y$ is the $N\times N$ orthogonal matrix with the eigenvectors $y_{1}%
,y_{2},\cdots,y_{N}$ in the columns obeying $Y^{T}Y=YY^{T}=I$. Since
$\beta_{1}=0$ and $y_{1}=\frac{u}{\sqrt{N}}$, it holds for $k>1$ that
$u^{T}y_{k}=0$, which implies that the sum of the components of eigenvector
$y_{k}$ for $k>1$ is zero (just as for any weighted Laplacian \cite{van2017pseudoinverse}). The position vector in (\ref{solution_positionvector}) is
rewritten as
\[
x[k]=Y\text{diag}(1+\beta)^{k}Y^{T}x[0]=\sum_{j=1}^{N}(1+\beta_{j})^{k}%
y_{j}\left(  y_{j}^{T}x[0]\right)
\]
Hence, we arrive at
\begin{equation}
	x[k]-\frac{u^{T}x[0]}{\sqrt{N}}u=\sum_{j=2}^{N}(1+\beta_{j})^{k}(y_{j}%
	^{T}x[0])\;y_{j}\label{solution_positionvector_in_terms_eigenvalues}%
\end{equation}
As explained above, the left-hand side is a translated position vector and
physically not decisive for the clustering process. Since $-1<\beta_{j}<0$ for
$j>1$, relation (\ref{solution_positionvector_in_terms_eigenvalues}) indicates
that, for $k\rightarrow\infty$, the right-hand side tends to zero and the
steady-state solution is clearly uninteresting for the clustering process. We
rewrite (\ref{solution_positionvector_in_terms_eigenvalues}) as
\[x[k]-\frac{u^{T}x[0]}{\sqrt{N}}u  = (1+\beta_{2})^{k}\bigg(  (y_{2}^{T}x[0])\;y_{2}+ \sum_{j=3}^{N}\left(  \frac{1+\beta_{j}}{1+\beta_{2}}\right)^{k}(y_{j}^{T}x[0])\;y_{j}\bigg).
\]
Since $|1+\beta_{2}|>|1+\beta_{3}|$, we observe that
\begin{equation}\label{position_vector_versus_y_2}
	\frac{x[k]-\frac{u^{T}x[0]}{\sqrt{N}}u}{(1+\beta_{2})^{k}\;(y_{2}^{T}   	x[0])}=y_{2}+O\left(  \frac{1+\beta_{3}}{1+\beta_{2}}\right)  ^{k},	
\end{equation}
which tells us that the left-hand side, which is a normalized or scaled,
shifted position vector, tends to the second eigenvector $y_{2}$ with an error
that exponentially decreases in $k$. Hence, for large enough $k$, but not too
large $k$, the scaled shifted position vector provides us the information on
which we will cluster the graph.

The steady state in Property \ref{prop_M_eigenvalues} can be regarded as a reference position of the nodes and does not affect the LCP process nor the $N\times 1$ eigenvector $y_2$, belonging to the second largest eigenvalue $(1 + \beta_2)$ of the $N\times N$ “operator” matrix $I + W - \text{diag}(W\cdot u)$, which is analogous to Fiedler clustering based on the $N\times N$ Laplacian $Q$. While the Laplacian matrix $Q$  essentially describes diffusion and not clustering, our operator $I + W - \text{diag}(W\cdot u)$ changes the nodal positions, based on attraction and repulsion, from which clustering naturally arises.

\subsection{Influence of $\alpha$ and $\delta$ on the eigenvalues $\beta_{k}$ and the eigenvector $y_2$}

\begin{property}
	\label{prop_alpha_delta}
	The two parameters in the matrix $W$ in (\ref{def_martrix_W}) satisfy the bounds \begin{eqnarray}
		0\leq\alpha &\leq&\frac{d_{\max}-1}{d_{\max}-\frac{1}{2}\left(  1+\frac{d_{\min}%
			}{d_{\max}}\right)  }\leq1 \label{bounds_alpha}\\
		0\leq\delta& \leq& \frac{1}{d_{\max}-\frac{1}{2}\left(  1+\frac{d_{\min}}{d_{\max
			}}\right)  } \label{bounds_delta}
	\end{eqnarray}
\end{property}
\textit{Proof:} Appendix \ref{sec_prop_alpha_delta}.

\begin{figure}[!h]
	\begin{center}
		\includegraphics[angle=0, scale=0.7]{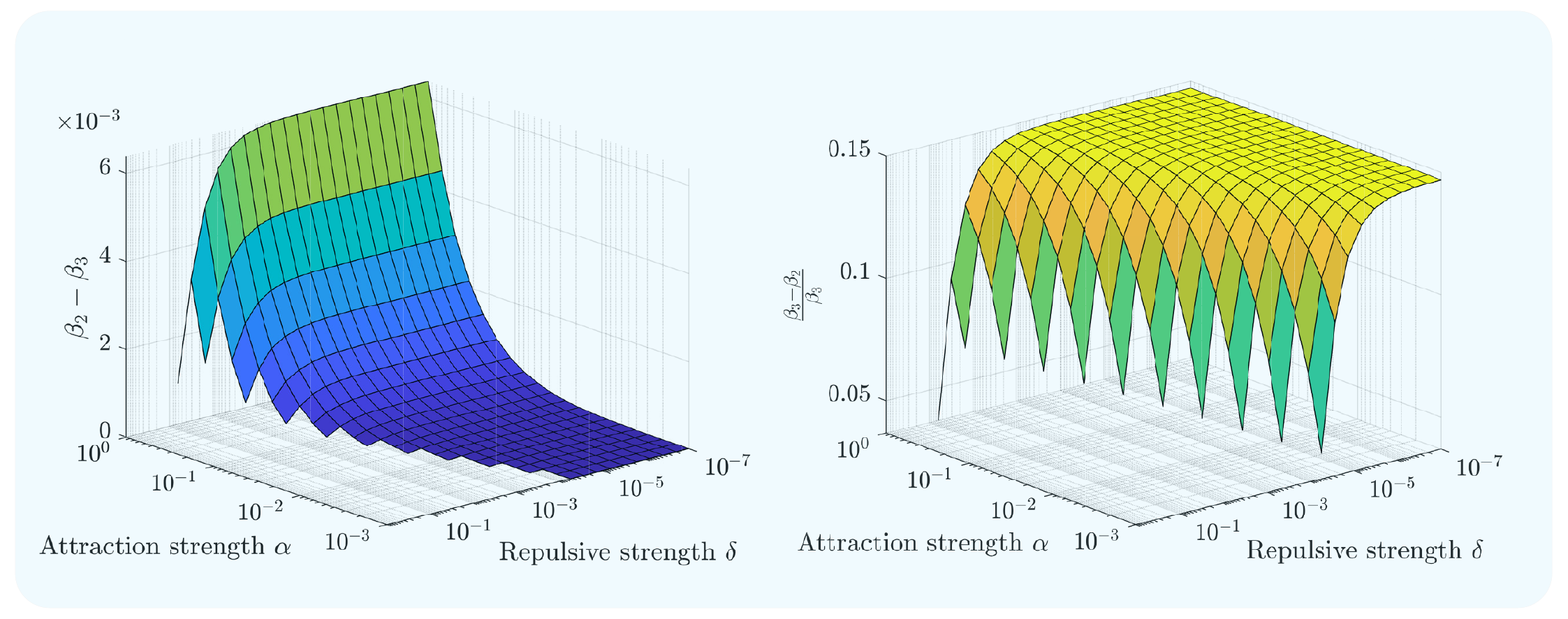}
		\caption{Gap $\beta_2 - \beta_3$ between the second and the third largest eigenvalue of the $N\times N$ matrix $W-\text{diag}\left(W\cdot u\right)$, for different values of the attractive $\alpha$ and repulsive $\delta$ strength (left figure). Relative difference $\frac{\beta_3 - \beta_2}{\beta_3}$, for different values of the attractive $\alpha$ and repulsive $\delta$ strength (right figure). An SSBM network with $N=1000$ nodes, $c=5$ clusters is used for both plots, with $b_{in} = 25$ and $b_{out}=2.5$.}
		\label{Fig_influence_eigenvaluegap}
	\end{center}
\end{figure}

Figure \ref{Fig_influence_eigenvaluegap} shows that influence of the attractive and repulsive strength $\alpha$ and $\delta$ on the eigenvalue gap $\beta_2-\beta_3$ is relatively small if $\alpha$ and $\delta$ are not too small and obeying the bounds (\ref{bounds_alpha}) and (\ref{bounds_delta}). While the difference increases when the attraction strength $\alpha$ is increasing, the repulsive strength $\delta$ has no visible influence on the eigenvalue gap.

\begin{figure}[!h]
	\begin{center}
		\includegraphics[angle=0, scale=0.75]{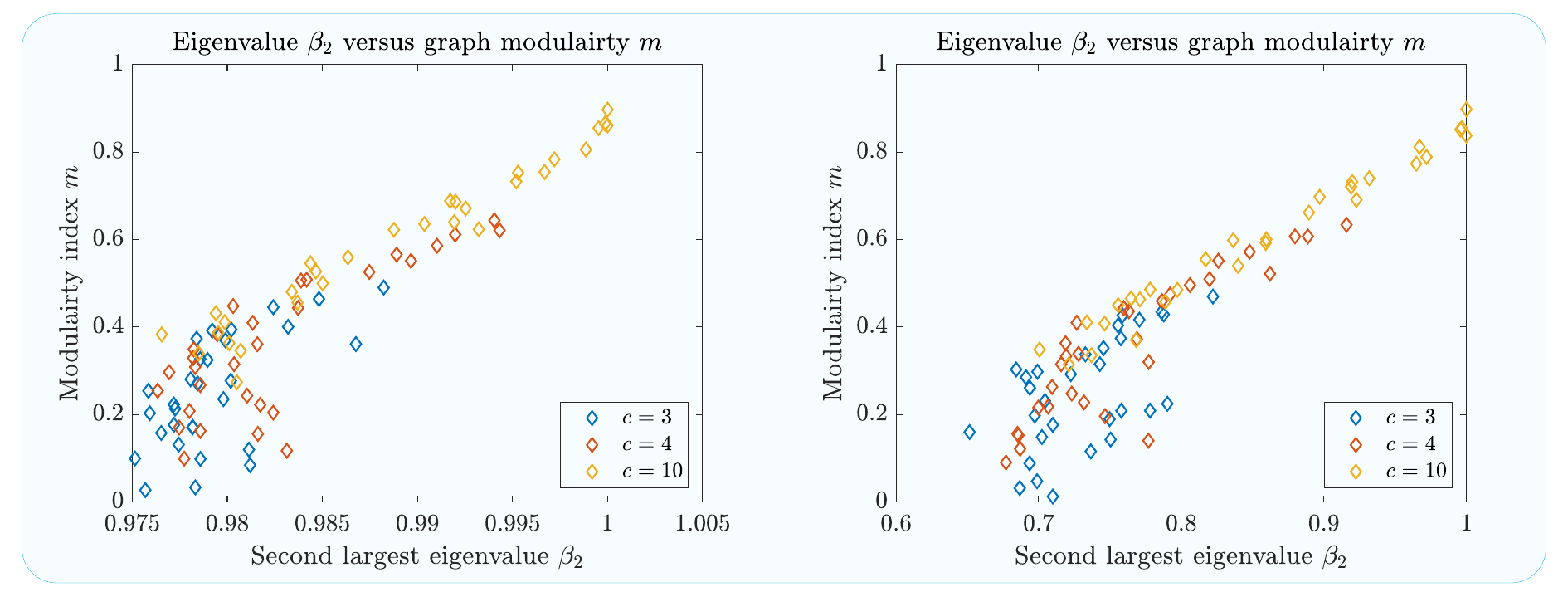}
		\caption{The eigenvalue $\beta_{2}$ versus the modularity index $m$ of an SSBM graph of $N=999$ nodes and $c=3$ clusters, and an SSBM graph of $N=1000$ nodes, with $c=4,10$ clusters, respectively. The parameters $b_{in}$ and $b_{out}$ are varied, while keeping average degree $d_{av}=7$ fixed. For each combination of $b_{in}$ and $b_{out}$, the modularity index $m$ and the eigenvalue $\beta_{2}$ are computed. The correlation is presented in case only interactions between direct neighbours are allowed (left figure) and in case interactions between each pair of nodes are allowed (right figure).}
		\label{Fig_beta_2_versus_modulairty}
	\end{center}
\end{figure}

The eigenvalue $\beta_{2}$ depends on the community structure of a graph. Figure \ref{Fig_beta_2_versus_modulairty} reveals positive correlation between the eigenvalue $\beta_{2}$ and the modularity index $m$ of a graph. As the modularity index increases, the eigenvalue $\beta_{2}$ approaches value $1$. In the limit case, when there are only intra-community links in the network, $\beta_{2}=1$, indicating the eigenvector $y_{2}$ represents a steady state.

\begin{figure}[!h]
	\begin{center}
		\includegraphics[angle=0, scale=0.6]{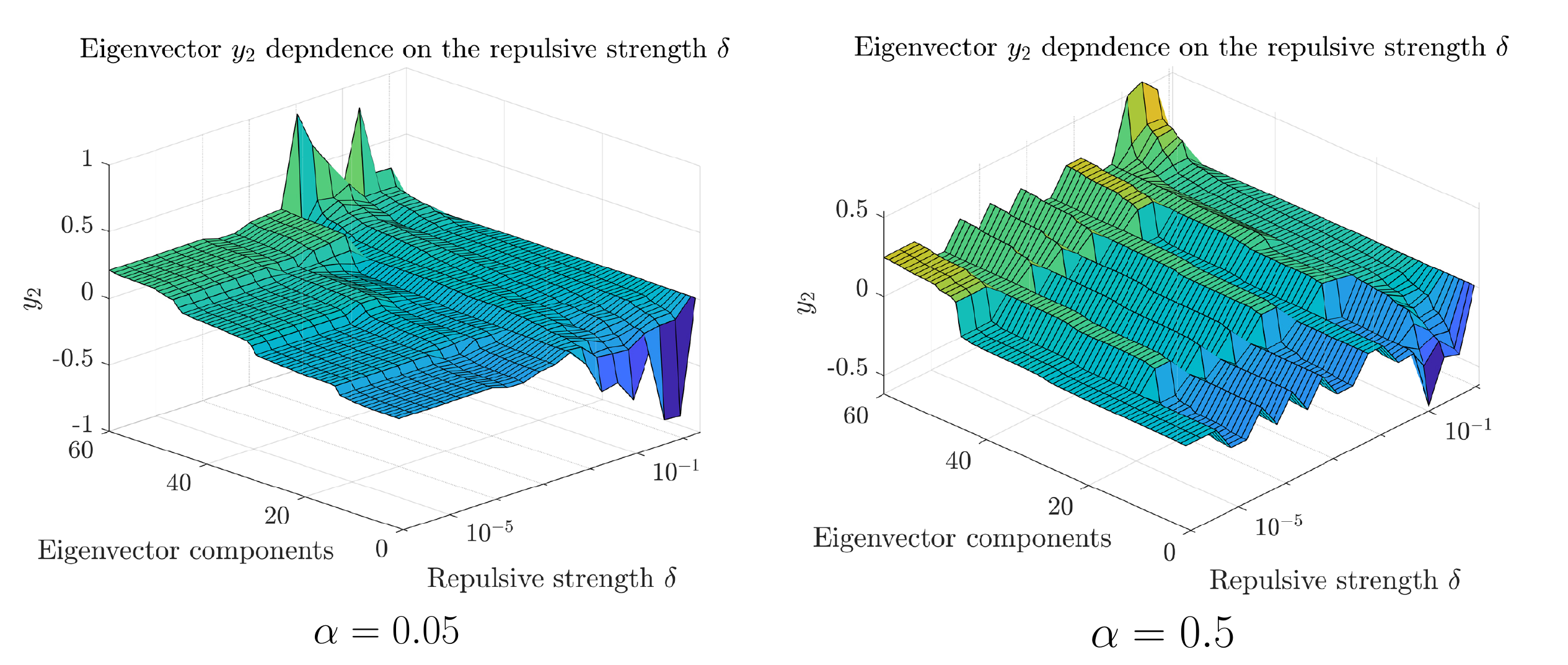}
		\caption{ Sorted eigenvector $\hat{y_{2}}$ components for different values of the repulsive strength $\delta$, in case of a SSBM network of $N=100$ nodes, $c=4$ clusters and with parameters $b_{in} = 25$ and $b_{out} = 1$. The attraction rate equals $\alpha = 0.05$ (left figure) and $\alpha = 0.5$ (right figure), while the repulsive strength $\delta$ obeys bounds in (\ref{bounds_delta}).}
		\label{Fig_influence_eigenvectory2}
	\end{center}
\end{figure}

Figure \ref{Fig_influence_eigenvectory2} reveals that the repulsive strength $\delta$ does not affect the eigenvector $y_2$ components significantly. Eigenvector $y_{2}$ components of nodes from the same cluster are better distinguished from the remaining components of $y_{2}$ for smaller values of repulsive strength $\delta$.

\section{From the eigenvector $y_2$ to clusters in the network}\label{Sec:Position_Vector_Analysis}

The interplay of the attractive and repulsive force between nodes drives the nodal position in discrete time $k$ eventually towards a steady state  $\lim\limits_{k\to \infty}x[k]=u$. However, the scaled and shifted position vector $x[k]$ in (\ref{position_vector_versus_y_2}) converges in time towards the second eigenvector $y_{2}$ with an exponentially decreasing error. In this section, we estimate the clusters in network, based on the eigenvector $y_{2}$.

By sorting the eigenvector $y_{2}$ to $\hat{y_2}$, the components of $y_{2}$ are reordered and the corresponding relabeling of the nodes of the network reveals a block diagonal structure of the adjacency matrix $A$. We define the $N\times N$ permutation matrix $R$ in a way the following equalities hold:
\begin{equation}\label{Eq_Permuation_Matrix_R}
	\begin{aligned}
		\hat{y_2} & = R\cdot y_2, \\ 
		\hat{(y_2)_i} = \left(y_{2}\right)_{r_{i}} &\leq \hat{(y_2)_j} = \left(y_{2}\right)_{r_{j}}, && i<j,
	\end{aligned}
\end{equation}
where the $N\times 1$ ranking vector $r = R\cdot w$ and $w = [1,2,\dots , N]$, with $r_{i}$ denoting the node $i$ ranking in the eigenvector $y_2$. The permutation matrix $R$ allow us to define the $N\times N$ relabeled adjacency matrix $\hat{A}$, the $N\times 1$ relabeled degree vector  $\hat{d}$ of $G$, and the $N\times 1$ sorted eigenvector $\hat{y_{2}}$ as follows:
\begin{equation}\label{Eq_Relabeled_A_and_d}
	\begin{cases}
		\hat{A} &= R^{T}\cdot A \cdot R \\  
		\hat{d} &= R \cdot d \\
		\hat{y_{2}} &= R \cdot y_2.
	\end{cases}
\end{equation}

\begin{figure*}[!]
	\begin{center}
		\includegraphics[angle=0, scale=0.7]{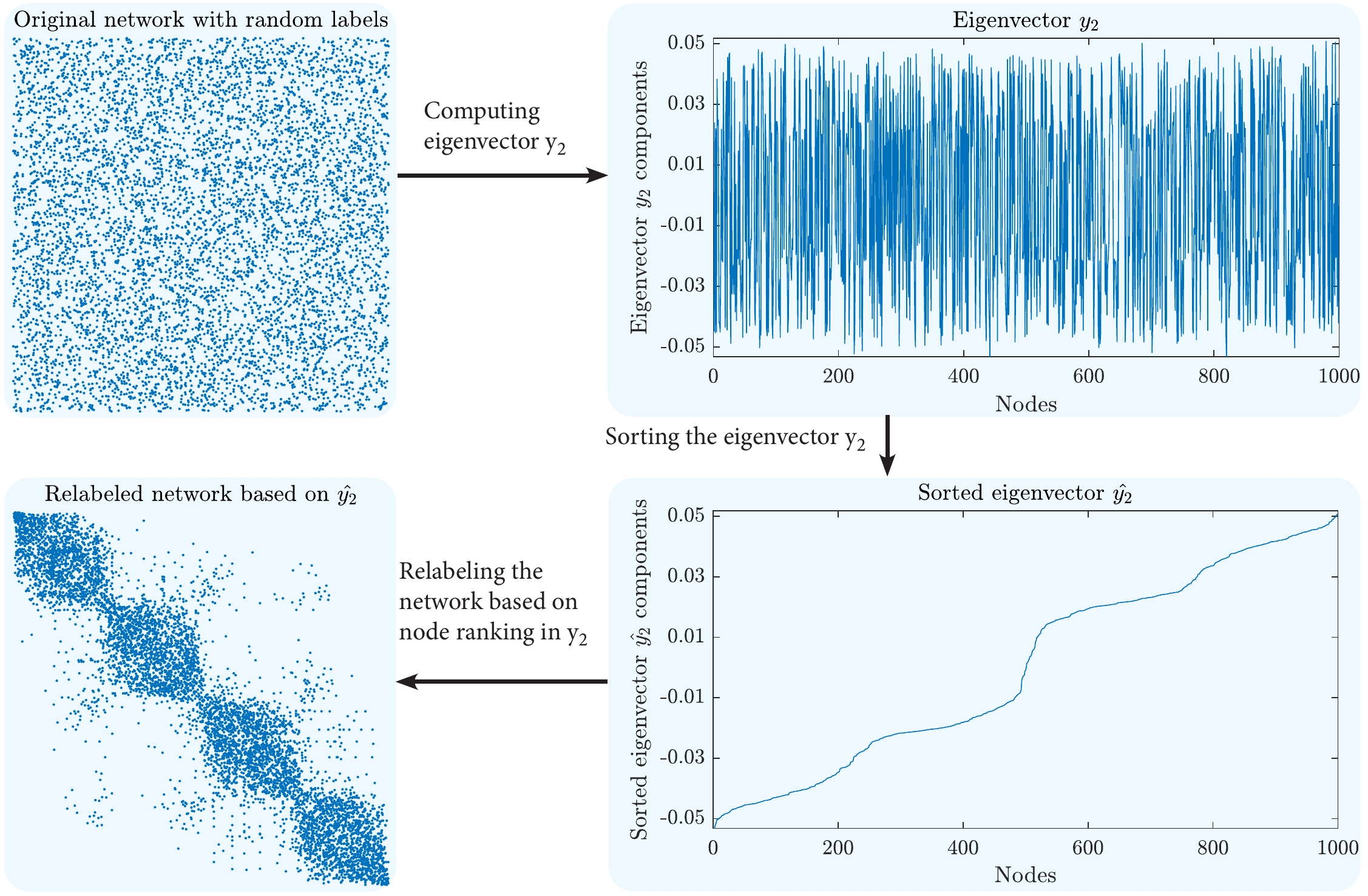}
		\caption{Adjacency matrix $A$ of an SSBM network of $N=1000$ nodes, $c=4$ clusters and parameters $b_{in}=26$, $b_{out}=0.67$ (top-left). Eigenvector $y_{2}$ components (top-right).	Sorted eigenvector $\hat{y_{2}}$ components (bottom-right). Relabeled adjacency matrix $\hat{A}$ based on the sorted eigenvector $\hat{y_{2}}$ (bottom-left).}
		\label{Fig_Model_Procedure_1}
	\end{center}
\end{figure*}

Groups of nodes that have relatively small difference in the eigenvector $y_{2}$ components, while relatively large difference compared to other nodes in the network, compose a cluster. Therefore, the community detection problem transforms into recognizing intervals of similar values in the sorted eigenvector $\hat{y_{2}}$.

Figure \ref{Fig_Model_Procedure_1} exemplifies the idea, where the adjacency matrix $A$ of a randomly labeled SSBM network of $N=1000$ nodes and $c=4$ clusters is presented in the upper-left part, as a heat map. The eigenvector $y_{2}$ is drawn in the upper-right part, while the sorted eigenvector $\hat{y_{2}}$ is drawn on the bottom-right side. Finally, the relabeled adjacency matrix $\hat{A}$, based on nodal ranking of $y_{2}$ is depicted on the lower-left side. The sorted eigenvector $\hat{y_{2}}$ reveals a stair with four segments, equivalent to four block matrices on the main diagonal in relabeled adjacency matrix $\hat{A}$.

The eigenvector $y_{2}$ represents a continuous measure of how similar neighbours of two nodes are. There are two different approaches to identify network communities for a given eigenvector $y_{2}$:
\begin{itemize}
	\item Cluster identification based on the sorted eigenvector $\hat{y_{2}}$. This approach is explained in subsection \ref{K_means_cluster_estimation}.
	\item Cluster identification based on the ranking vector $r$. This approach does not rely on the eigenvector $y_{2}$ components, but solely on nodal ranking, as explained in subsection \ref{Recursive_algorithm_sec}.
\end{itemize}

\subsection{Community detection based on nodal components of the eigenvector $y_{2}$}\label{K_means_cluster_estimation}
To identify clusters, we observe the difference in eigenvector $y_{2}$ components between nodes with adjacent ranking. If $\left(\hat{y_{2}}\right)_{i+1}-\left(\hat{y_{2}}\right)_{i} < \theta$, where  $\theta$ denotes a predefined threshold, then the nodes $r_{i}$ and $r_{i+1}$ belong to the same cluster, else the nodes $r_{i}$ and $r_{i+1}$ are boundaries of two adjacent clusters. The resulting cluster membership function is
\begin{equation}\label{Eq:Modularity_matrix_C:position}
	C_{r_{i+1},r_{i}}=	\begin{cases}
		1 & \text{$\left(\hat{y_{2}}\right)_{i+1}-\left(\hat{y_{2}}\right)_{i} < \theta$}\\
		0 & \text{otherwise},
	\end{cases}
\end{equation}
where the threshold value $\theta$ is determined heuristically, as provided in  section \ref{Sec_Results}. The cluster estimation in (\ref{Eq:Modularity_matrix_C:position}) can be improved by using other more advanced approaches, such as the K-means algorithm. 

\subsection{Modularity-based community detection}\label{Recursive_algorithm_sec}

By implementing (\ref{Modularity_matrix_C}) and (\ref{Eq_Relabeled_A_and_d}) into (\ref{Eq_modularity}) we obtain:
\begin{equation}\label{Eq_modularity_2}
	m =  \frac{1}{2L}\cdot u^{T}\cdot \left(\hat{A}\circ \hat{C} - \frac{1}{2L}\cdot \left(\hat{d}\cdot \hat{d}^{T}\right)\circ \hat{C}\right)\cdot  u,
\end{equation}
where $\hat{C} = R^{T}\cdot C \cdot R$. As shown in Figure \ref{Fig_Model_Procedure_1}, the network relabeling based on the ranking vector $r$ reveals block diagonal structure in $\hat{A}$. Thus, the relabeled modularity matrix $\hat{C}$ has the following block diagonal structure:
\begin{equation}\label{Eq_Modularity_Matrix_Hat_C}
	\hat{C}=\begin{bmatrix}
		J_{n_{1}\times n_{1}} & O_{n_{1}\times n_{2}} & \dots & O_{n_{1}\times n_{c}} \\
		O_{n_{2}\times n_{1}} & J_{n_{2}\times n_{2}} & \dots & O_{n_{1}\times n_{c}} \\
		\vdots & \vdots & \dots & \vdots \\
		O_{n_{1}\times n_{1}} & O_{n_{c}\times n_{2}} & \dots & J_{n_{c}\times n_{c}} \\
	\end{bmatrix},
\end{equation}
where $c$ denotes number of clusters in network, where the $i$-th cluster is composed of $n_{i}$ nodes. We define the $N\times 1$ vectors $\hat{e_{i}}$ for $i=\{1,2,\dots ,c\}$ as
\begin{equation}\label{Eq_vectors_e_i}
	\hat{e_{i}}=\begin{bmatrix}
		O_{\left(1 \times \sum_{j=1}^{i-1}n_{j}\right)} &
		u_{\left(1 \times n_{i}\right)} &	O_{\left(1\times \sum_{j=i+1}^{N}n_{j}\right)} \end{bmatrix}^{T},
\end{equation}
that allows us to redefine $\hat{C} = \sum_{i=1}^{c}\hat{e_{i}}\cdot \hat{e_{i}}^{T}$ and further simplify (\ref{Eq_modularity_2}):
\begin{equation}\label{Eq_Modulaity_m_with_E_vectors}
	m =  \frac{1}{2L}\cdot \sum\limits_{i=1}^{c}\hat{e_{i}}^{T}\cdot \left(\hat{A}-\frac{1}{2L}\cdot \left(\hat{d}\cdot \hat{d}^{T}\right)\right)\cdot \hat{e_{i}}.
\end{equation}
Since the vector $\hat{e_{i}}$ consists of zeros and ones, the equation (\ref{Eq_Modulaity_m_with_E_vectors}) represents the sum of elements of the matrix $\left(\hat{A}-\frac{1}{2L}\cdot \left(\hat{d}\cdot \hat{d}^{T}\right)\right)$ corresponding to each individual cluster. 

\begin{algorithm}[H]
	\caption{Recursive algorithm for cluster estimation
		\label{alg:cluster-estimation}}
	\begin{algorithmic}[1]
		\Require{$\hat{A}$ and $\hat{d}$ are the relabeled adjacency matrix $A$ and the degree vector $d$ (\ref{Eq_Relabeled_A_and_d}), while $L$ denotes number of links. The modularity threshold is denoted by $\theta$. The function returns the $c\times 1$ vector $b$, whose elements are cluster borders in a relabeled graph.}
		\Statex
		\Function{EstimateClusters}{$\hat{A},\, \hat{d},\, N,\, L,\, \theta$}
		\State $d_{f},d_{b},p,q \gets O_{N\times 1}$
		\State $\left(d_{f}\right)_{1} \gets \hat{d}_{1}$
		\State $\left(d_{b}\right)_{N} \gets \hat{d}_{N}$
		\State $p_{1} \gets -\frac{\hat{d}_{1}^{2}}{(2L)^{2}}$
		\State $q_{N} \gets -\frac{\hat{d}_{N}^{2}}{(2L)^{2}}$
		\For{$i \gets 2 \textrm{ to } N$} 
		\State $l \gets N - i$
		\State $\left(d_{f}\right)_{i} \gets \left(d_{f}\right)_{i-1} + \hat{d}_{i}$
		\State $\left(d_{b}\right)_{N-i+1} \gets \left(d_{b}\right)_{N-i+2} + \hat{d}_{N-i+1}$
		\State $ s \gets \frac{ \sum_{j=1}^{i}\hat{a}_{ij}}{L} - \frac{	2\cdot \hat{d}_{i}\cdot (d_{f})_{i-1} + \hat{d}_{i}^{2}}{(2L)^{2}}$
		\State $ t \gets \frac{\sum_{j=1}^{i}\hat{a}_{N-j+1,l+1}}{L} - \frac{2\cdot \hat{d}_{l+1}\cdot (d_{b})_{l+2} + \hat{d}_{l+1}^{2}}{(2L)^{2}} $ 
		\State $ p_{i} \gets p_{i-1} + s$
		\State $q_{l+1} \gets q_{l+2} + t$ 
		\EndFor
		\State $r \gets \operatorname*{arg\,max}_{\mathcal{N}} \left(p+q\right)$
		\If{$(p+q)_{r} > \theta$}
		\State $\hat{A_{1}},\, \hat{d_{1}},\,N_{1} \gets$ sub-matrix(vector) corresponding to the first cluster $\{1,2, \dots , r\}$
		\State $\hat{A_{2}},\, \hat{d_{2}},\,N_{2} \gets$ sub-matrix(vector) corresponding to the second cluster $\{r+1,r+2, \dots , N\}$
		\State \Return $\hat{b} \gets \begin{bmatrix}
			\text{EstimateClusters(}\hat{A_{1}},\hat{d_{1}},N_{1},L,p_{r}\text{)} \\
			r, \\
			\text{EstimateClusters(}\hat{A_{2}},\hat{d_{2}},N_{2},L,q_{r}\text{)}
		\end{bmatrix}$
		\Else
		\State \Return{$\hat{b} \gets\emptyset$}
		\EndIf
		\EndFunction
	\end{algorithmic}
\end{algorithm}

We estimate clusters for a given ranking vector $r$ by optimising the modularity $m$ recursively.
In the first iteration, we examine all possible partitions of the network in two clusters and compute their modularity. The partition that generates the highest modularity is chosen.
In the second iteration, we repeat for each subgraph the same procedure and find the best partitions into two clusters. Once we determine the best partitions for both subgraphs, we adopt them if the obtained modularity of the generated partition exceeds the modularity of a parent cluster from the previous iteration. The recursive procedure stops when the modularity $m$ cannot be further improved, as described by pseudocode (\ref{alg:cluster-estimation}). This version of the proposed process is denoted as LCP in section \ref{Sec_Results}.

\subsection{Modularity-based community detection for a known number of communities}\label{Recursive_algorithm_given_N_sec}

The algorithm \ref{alg:cluster-estimation} also applies for graph partition with a known number of communities $c$. In that case, instead of stopping the recursive procedure described in algorithm \ref{alg:cluster-estimation} when the modularity $m$ cannot be further improved, we stop at iteration $(\log_{2}c+1)$. In each iteration, the partition with the maximum modularity is accepted, even if negative.

As a result, we obtain $2c$ estimated clusters with the $2c\times 2c$ aggregated modularity matrix $M_{c}$:
\begin{equation}\label{Eq_Modularity_Matrix_M_c}
	\left(M_{c}\right)_{gh} = \sum\limits_{i\in g,j\in h}\left(\hat{A}-\frac{1}{2L}\cdot \hat{d}\cdot \hat{d}^{T}\right)_{ij},
\end{equation}
where $g,h \in \{1,2, \dots , 2c\}$ denote estimated communities. The aggregated modularity matrix $M_{c}$ allows us to merge adjacent clusters, until we reach $c$ communities in an iterative way. We observe the $(2c-1\times 1)$ vector $\mu$, where $\mu_{g} = \left(M_{c}\right)_{g,g+1}$. The maximum element of $\mu$ indicates which two adjacent clusters can be merged, so that modularity index $m$ is negatively affected the least.  By repeating this procedure $c$ times, we end up with the graph partition in $c$ clusters. This version of the proposed process is denoted as LCP$_{n}$ in Section \ref{Sec_Results}.

\subsection{Non-back tracking method versus LCP}\label{Non_back_tracking_as_a_process}

Angel \textit{et al.} \cite[p.12]{angel2015non} noted that the $2N$ non-trivial eigenvalues of the $2L\times 2L$ non-back tracking matrix $B$ from (\ref{Eq:Non-back tracking B})  are contained in eigenvalues of the $2N\times 2N$ matrix $B^{*}$:
\begin{equation}\label{Eq:Non_back_tracking_2N}
	B^{*} = \begin{bmatrix}
		A & I - \Delta \\
		I & O
	\end{bmatrix},
\end{equation}
where the $N\times N$ matrix with all zeros is denoted as $O$. The $2N\times 2N$ matrix $B^{*}$, written as
\[B^{*} = \begin{bmatrix}
	I + \left(A - \Delta\right) + \left(\Delta - I\right) & -\left(\Delta - I\right) \\
	I & O
\end{bmatrix}\]
can be considered as a state-space matrix of a process on a network, similar to our LCP process we in (\ref{clustering_law_discrete_time}), with the last $N$ states storing delayed values of the first $N$ states. The $2N\times 2N$ matrix $B^{*}$ defines the set of $N$ second-order difference equations, where the governing equation for the node $i$ position is
\begin{eqnarray}
	x_{i}[k+1] = x_{i}[k] + \sum\limits_{j\in \mathcal{N}_{i}}(x_{j}[k]-x_{i}[k])\nonumber+ \left(d_{i}-1\right)\cdot \left(x_{i}[k]-x_{i}[k-1]\right) \label{Eq_nbtm_process}
\end{eqnarray}
We recognize the second term in (\ref{Eq_nbtm_process}) as an attraction force between neighbouring nodes with uniform intensity, while in our LCP (\ref{clustering_law_discrete_time}) the attraction force intensity is proportional to the number of neighbours two adjacent nodes share. Further, while we propose a repulsive force between adjacent nodes in (\ref{clustering_law_discrete_time}), node $i$ in (\ref{Eq_nbtm_process}) is repulsed from its previous position $x_{i}[k]$ in direction of the last position change $(x_{i}[k]-x_{i}[k-1])$.

We implement the weighted intensity of the attractive force as in (\ref{clustering_law_discrete_time}), ignoring the repulsive force by letting $\delta = 0$, and define the $2N\times 2N$ matrix $W^{*}$, corresponding to $B^{*}$,
\begin{equation}\label{Eq_Non_back_tracking_Matix_W}
		W^{*} = \begin{bmatrix}
			I + \alpha\cdot \Big(A\circ A^{2} +A - \text{diag}\left(\left(A\circ A^{2} + A\right)\cdot u\right)\Big) + \left(\Delta - I\right) & -\left(\Delta - I\right) \\
			I & O
		\end{bmatrix}.
\end{equation}
We estimate the number of clusters $c$ in a network from $W^{*}$ similarly as in the non-back tracking method in Sec. \ref{Sec_non_back_tracking_matrix} by counting the number of eigenvalues in $W^{*}$ with real component larger than $\sqrt{\lambda_1(W^{*})}$. This approach is denoted as LCP$_{c}$ in Section \ref{Sec_Results}.

\section{Reducing intensity of forces between clusters}\label{Sec_Iterative_Computation}

The idea behind a group of methods in community detection, called divisive algorithms, consists of determining the links between nodes from different clusters. Once these links have been identified, they are removed and thus only the intra-community links remain \cite{girvan2002community}. 
We invoke a similar idea to our linear clustering process.

An outstanding property of our approach is that the LCP defines the nodal position as a metric, allowing us to perform clustering in multiple ways. The position distance between any two, not necessarily adjacent nodes indicates how likely the two nodes belong to the same cluster. Then, the position metric also allows us to classify links as either intra- or inter-community.
Thus, we iterate the linear clustering process (\ref{clustering_law_discrete_time}) and, in each iteration, we identify and scale the weights of the inter-community links. 

The attraction and repulsive forces are defined as linear functions of the position difference between two neighbouring nodes, as presented in Figure \ref{Fig_Forces_1}. While linear functions greatly simplify the complexity and enable a rigorous analysis, the linearity of forces introduces some difficulties in the process. Firstly, as two adjacent nodes are further away, both the attractive and the repulsive force between them increase in intensity. Similarly, as the neighbouring nodes are closer on a line, both forces decrease in intensity and converge to zero as the nodes converge to the same position. Secondly, the attractive force between any two neighbouring nodes is always of higher intensity than the repulsive force, causing the process to converge towards the trivial steady-state. 

Non-linearity in the forces can be introduced in the proposed linear clustering process iteratively by scaling the weights of inter-community links between iterations, that artificially decreases the strength of forces between the two nodes from different clusters. In other words, we reduce the importance of links between nodes from different clusters, based on the partition from previous iteration. 

\subsection{Scaling the weights of inter-community links}\label{removing_links_sub_section}
The difference $|\left(y_{2}\right)_{i} - \left(y_{2}\right)_{j}|$ in the eigenvector $y_{2}$ components of nodes $i$ and $j$ indicates how similar neighbourhoods of these nodes are. A normalized measure for the difference in neighbouring nodes $i$ and $j$ is the difference $\left(|r_{i} - r_{j}|\right)$ of their rankings in the sorted eigenvector $\hat{y_{2}}$. Thus, links that connect nodes with the highest ranking difference are most likely inter-community links.
We define the $N\times N$ scaling matrix $S$ as follows:
\begin{equation}\label{Ranking_Distance_eq}
	s_{ij} =\begin{cases}
		1, & \text{if } \lvert r_{j} - r_{i}\rvert < \theta_{r} \\
		\gamma, & \text{otherwise },
	\end{cases}
\end{equation}
where the $ij$-th element equals 1 if the absolute value of the ranking difference between nodes $i$ and $j$ is below a threshold $\theta_{r}$, otherwise some positive value $0\le \gamma \le 1$. Based on the $N\times N$ scaling matrix $S$ in (\ref{Ranking_Distance_eq}), we update the governing equation as follows:
\[
x[k+1] = \left(  I + \tilde{W} - \text{diag}\left( \tilde{W} \cdot u \right)  \right)  \cdot
x[k],
\]
where $\tilde{W} = S\circ W$. 
\begin{figure*}[!]
	\begin{center}
		\includegraphics[angle=0, scale=0.85]{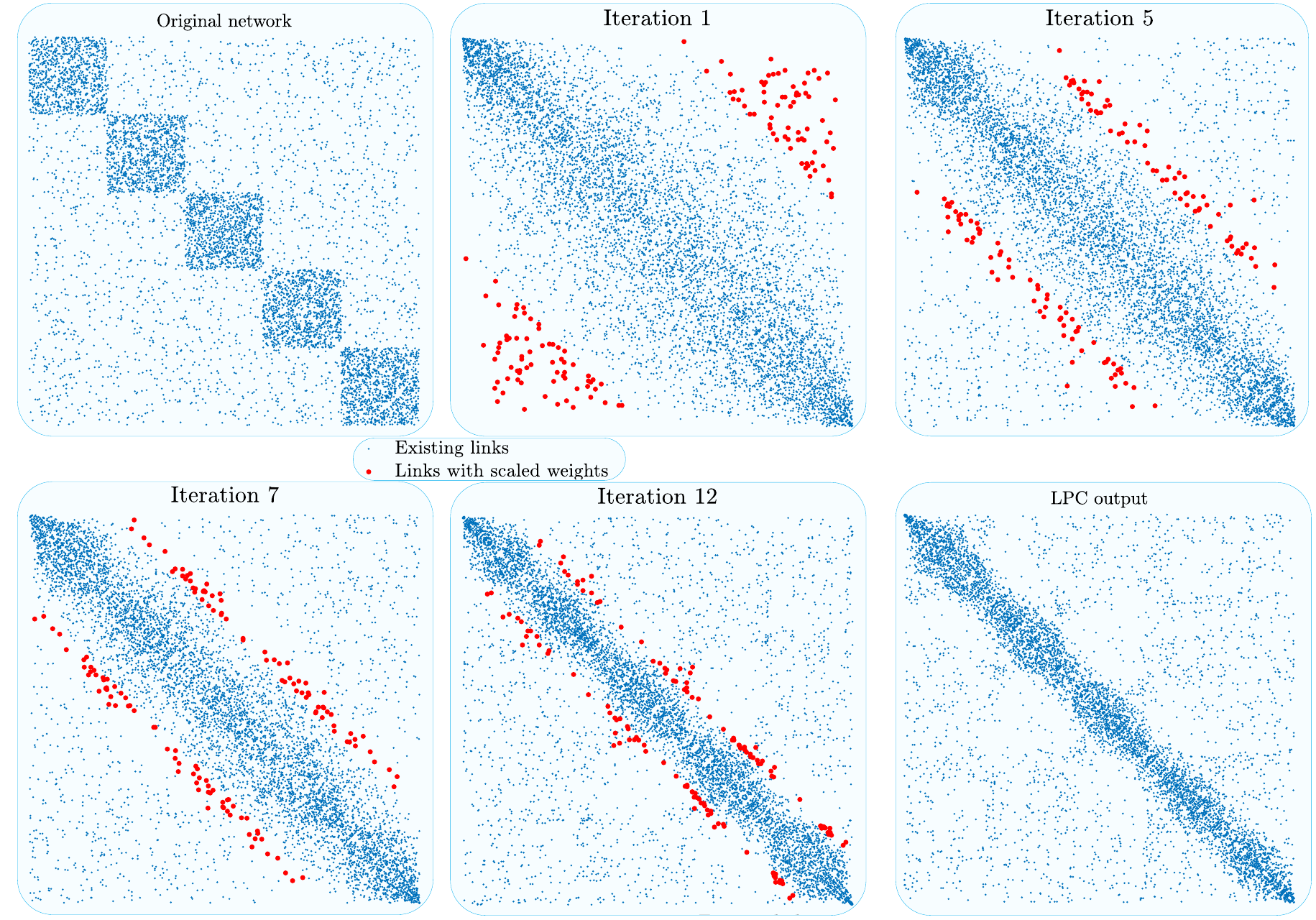}
		\caption{Adjacency matrix $A$ of an SSBM network of $N=1000$ nodes, $c=5$ clusters of equal size, with parameters $b_{in}=26$ and $b_{out}=2.25$  (top-left). Following 4 subfigures present the relabeled adjacency matrix based on the ranking vector $r$ in iterations $1,5,7$ and $12$, respectively. In each iteration, the weights of $2\%$ links are scaled (red colour). The weight of each link is allowed to be scaled once. The relabelled adjacency matrix $\hat{A}$ after $15$ iterations of scaling weights of links between clusters (bottom-right).}
		\label{Removing_Links_Fig}
	\end{center}
\end{figure*}
Scaling the link weights in (\ref{Ranking_Distance_eq}) only impacts the clustering process in (\ref{clustering_law_matrix_equation}), as defined in the equation above. However, modularity-based community detection, explained in Section \ref{Recursive_algorithm_sec}, operates on the $N\times N$ adjacency matrix $A$ in each iteration. Therefore, our implementation of scaling the weights of inter-community connections in network helps the process to better distinguish between clusters (i.e. eventually provides better relabeling in (\ref{Eq_Relabeled_A_and_d})), without modifying the $N\times N$ adjacency matrix $A$ and, hence, without negatively affecting the modularity $m$ optimisation in Algorithm \ref{alg:cluster-estimation}.
An example of removing links (i.e. $\gamma = 0$) is depicted on Figure \ref{Removing_Links_Fig}, where in each iteration weights of $\frac{15}{4}\%$ identified inter-cluster links are scaled. Scaling the weights of links between clusters significantly improves the quality of the identified graph partition.

\section{Benchmarking LCP with other clustering methods}\label{Sec_Results}

In this section, we benchmark the linear clustering  process (\ref{clustering_law_discrete_time}) against popular clustering algorithms in Appendix \ref{App_Clustering_Algorithms} on random networks.

\subsection{Complexity of LCP}\label{SubSec:Complexity}

The computational complexity of LCP consists of three parts: the computation of (i) the $N\times N$ matrix $W$ in (\ref{def_martrix_W}), (ii) the $N\times 1$ eigenvector $y_{2}$ of the matrix $W-$diag$(Wu)$ and (iii) the identification of the clusters based on the sorted eigenvector $\hat{y_{2}}$. 

\subsubsection{Computing the $N\times N$ matrix $W$}\label{SubSubSec:A^2}
The $N\times N$ matrix $A\circ A^{2}$ in (\ref{def_martrix_W}) requires the highest computational effort. Generally, computing the square of a matrix involves $O(N^{3})$ elementary operation, but the zero-one structure of the adjacency matrix significantly reduces the operations. We provide below an efficient algorithm for the computation of $A\circ A^{2}$, whose entries determine the number of $2$-hop walks between any two direct neighbours in the network. 

\begin{algorithm}[H]
	\caption{Computation of the $N\times N$ matrix $A\circ A^{2}$
		\label{alg:adjacency-squared}.}
	\begin{algorithmic}[1]
		\Require{$A$ denotes the adjacency matrix, $N$ denotes number of links, while the set of node $i$ neighbours is denoted by $\mathcal{N}_{i}$.}
		\State $A_{s} \gets O_{N\times N}$
		\For{$i \gets 1 \textrm{ to } N$}
		\For{$j \gets \mathcal{N}_{i}$}
		\For{$m \gets \left(\mathcal{N}_{j}\setminus \{1,2,\dots ,i\}\right)\cap \mathcal{N}_{i}$}
		\State $\left(A_{s}\right)_{i,m} \gets  \left(A_{s}\right)_{i,m} + 1$	\Comment{Account for the 2-hop walk $(i\rightarrow j \rightarrow m)$}
		\EndFor
		\EndFor
		\EndFor
		\State $A_{s} \gets A_{s} + A_{s}^{T}$
		\State \Return{$A_{s}$}
	\end{algorithmic}
\end{algorithm}

\begin{figure}[!h]
	\begin{center}
		\includegraphics[angle=0, scale=0.82]{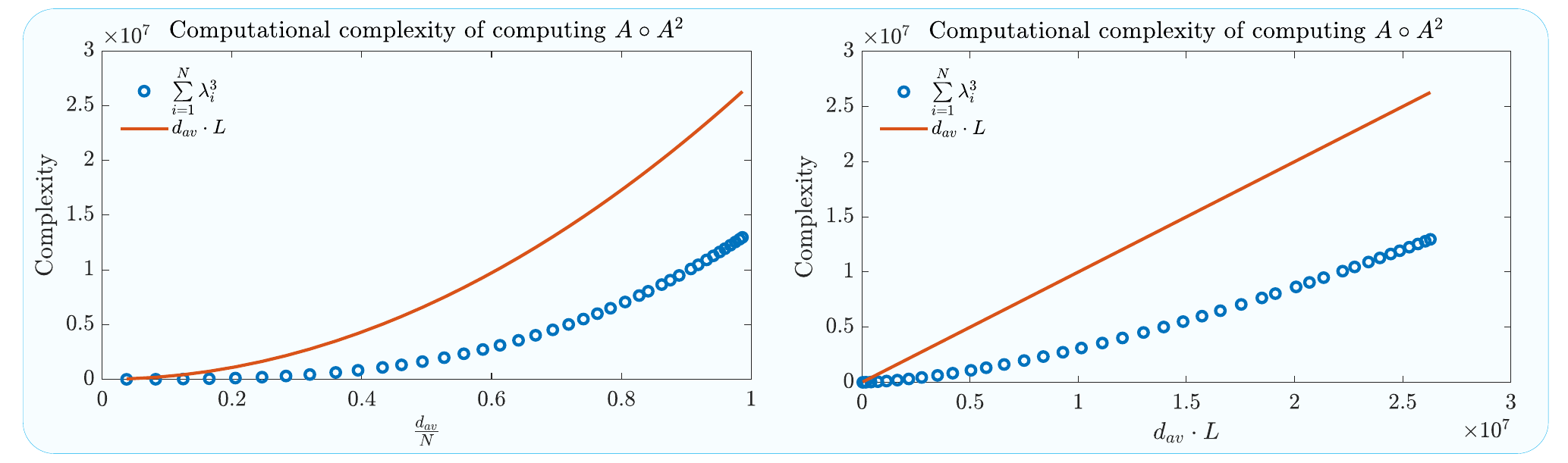}
		\caption{Sum of the cubed eigenvalues $\lambda$ of the adjacency matrix $A$ (blue circles) and product of the average degree $d_{av}$ and number of links $L$ (red line), for an Erdős–Rényi random graph with $N=300$ nodes, versus the relative mean degree $\frac{d_{av}}{N}$ (left-figure) and $d_{av}\cdot L$ (right-figure).}
		\label{Squared_A_Fig}
	\end{center}
\end{figure}

We initialize the $N\times N$ matrix $A\circ A^{2}$ with zeros and only compute elements above the main diagonal, because $A\circ A^{2}$ is symmetric. The algorithm identifies all 2-hop walks between any two direct neighbours and accordingly updates the matrix. Let us consider a node $i$ with $d_{i}$ neighbours, denoted as $\mathcal{N}_{i}$. For a neighbouring node $ j\in \mathcal{N}_{i}$, we increment the elements $\left(A\circ A^{2}\right)_{im}$ by 1, where $m\in \left(\mathcal{N}_{j}\setminus \{1,2,\dots ,i\}\right)\cap \mathcal{N}_{i}$, accounting for 2-hop walks $i\rightarrow j \rightarrow m$. By repeating the procedure for each node, we compute all the elements above the main diagonal. Finally, we sum the generated matrix with its transpose to obtain $A\circ A^{2}$. Since the algorithm \ref{alg:adjacency-squared} is based on incrementing the matrix entries per each 2-hop walk between direct neighbours, the number of operations equals the sum $s = u^{T}\cdot \left(A\circ A^{2}\right)\cdot u$ of all elements of $A\circ A^{2}$
\begin{equation}\label{Eq:A_squared_complexity}
	s =  \sum\limits_{i=1}^{N}\sum\limits_{j=1}^{N} \lambda_{i}\cdot \lambda_{j}^{2}\cdot
	u^{T}\left(x_{i}\circ x_{j}\right)\cdot \left(x_{i}\circ x_{j}\right)^{T} u
\end{equation}
The eigenvectors of the adjacency matrix $A$ are orthogonal. Therefore $\left(x_{i}\circ x_{j}\right)^{T}\cdot u = x_{i}^{T}\cdot x_{j} = 0$ if $i\ne j$, otherwise it equals 1 and (\ref{Eq:A_squared_complexity}) further simplifies to
\begin{equation}\label{Eq:A_squared_complexity_solution}
	s = \sum\limits_{i=1}^{N}\lambda_{i}^{3},
\end{equation}
which equals 6 times number of triangles in the network \cite[p. 31]{van2010graph}, because a 2-hop walk between adjacent nodes $i$ and $j$ over a common neighbour $m$ is equivalent to a triangle $i \rightarrow m \rightarrow j \rightarrow i$. The computational complexity of $A\circ A^{2}$ thus reduces to $O\left(d_{av}\cdot L\right)$, as presented in Figure \ref{Squared_A_Fig}. For a given matrix $A\circ A^{2}$, the computational complexity of the $N\times N$ matrix $W$ is $O(L)$, because (\ref{def_martrix_W}) can be transformed into Hadamard product terms (i.e. element-based operations).

\subsubsection{Computing the $N\times 1$ eigenvector $y_{2}$}\label{SubSubSec:y_2}
The eigenvector $y_{2}$ corresponds to the second largest eigenvalue $\beta_{2}$ of the $N\times N$ matrix $W-\text{diag}(W\cdot u)$. The largest eigenvalue $\beta_{1}=1$ corresponds to the eigenvector $y_{1}=\frac{1}{\sqrt{N}}u$. Computing the eigenvector $y_{2}$ is equivalent to computing the largest eigenvector of the matrix $W-\text{diag}(W\cdot u)-\frac{1}{N}\cdot u\cdot u^{T}$, which can be executed using the power method \cite{van2010graph}, for a given matrix $W$, with computational complexity $O\left(L\right)$.

\subsubsection{Computing the cluster membership function}\label{SubSubSec:C}
We apply the recursive algorithm \ref{alg:cluster-estimation} to identify communities based on the $N\times 1$ eigenvector $y_{2}$. The number of iterations of the algorithm ideally equals $T=\log_{2}c$, while in worst case scenario there are $c$ iterations.
Given a fixed number $c$ of communities, the computational complexity within an iteration is $O\left(L\right)$, as shown in pseudocode \ref{alg:cluster-estimation}. The number of clusters $c$ may depend upon $N$ and is in worst case equal to $N$. Thus, computational complexity increases in worst case to $O(N\cdot L)$.

\subsubsection{Scaling the inter-community links}\label{SubSubSec:Scaling_Links}
Between two iterations of the linear clustering process, we identify inter-community links and scale their weights, as defined in (\ref{Ranking_Distance_eq}). The computational complexity of this step is $O\left(L\right)$, as the ranking difference of neighbouring nodes is computer over each link.

Finally, computational complexity of the entire proposed clustering process equals $O(N\cdot L)$, because $d_{av} = O(N)$.

\subsection{Clustering performances on stochastic block generated graphs }\label{SubSec:SBMM_Results}

\begin{figure*}[!]
	\begin{center}
		\includegraphics[angle=0, scale=1.00]{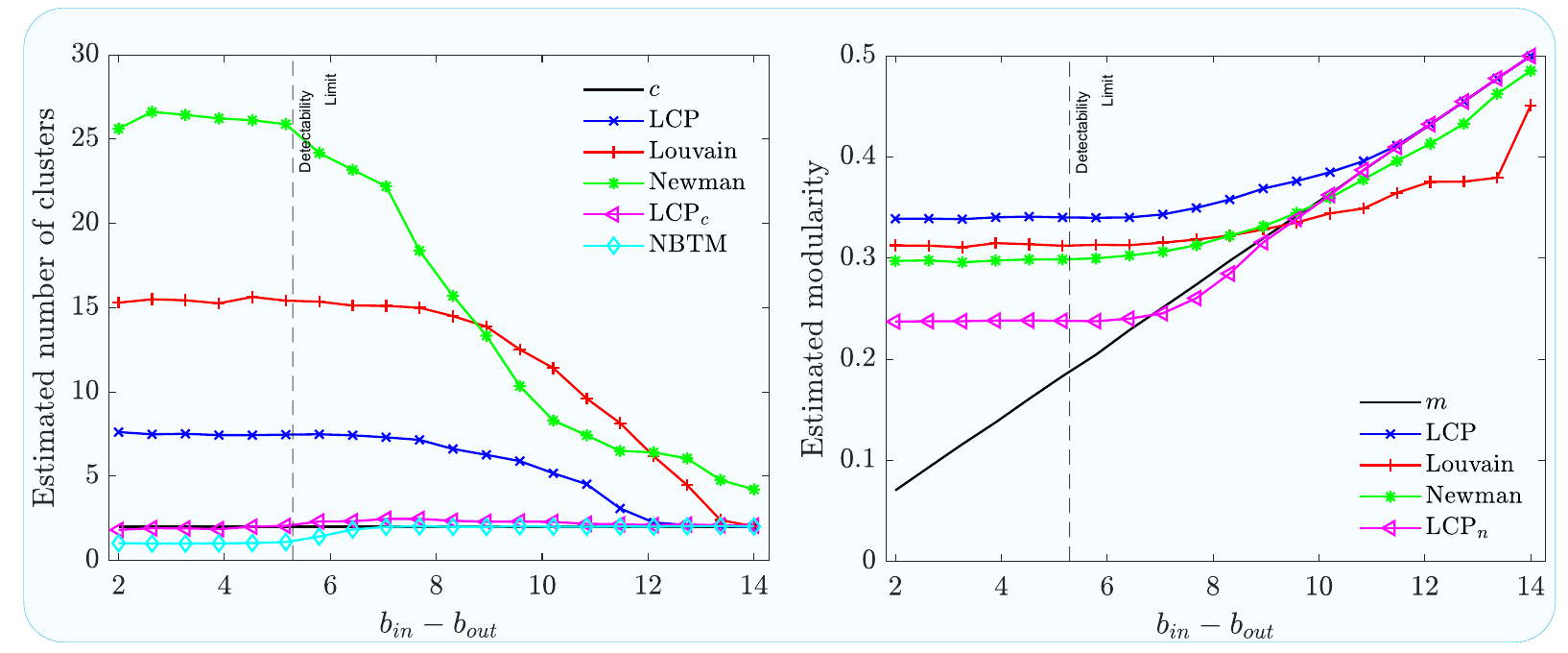}
		\caption{ The estimated number of clusters (left-figure) in SSBM graphs with $N=1000$ nodes, average degree $d_{av}=7$ and $c=2$ clusters, for different values of parameters $b_{in}$ and $b_{out}$. The modularity of the estimated partitions is presented in right figure. The vertical dashed line indicates the clustering detectability threshold.}
		\label{Fig_Results_2_clusters}
	\end{center}
\end{figure*}

\begin{figure*}[!]
	\begin{center}
		\includegraphics[angle=0, scale=1.00]{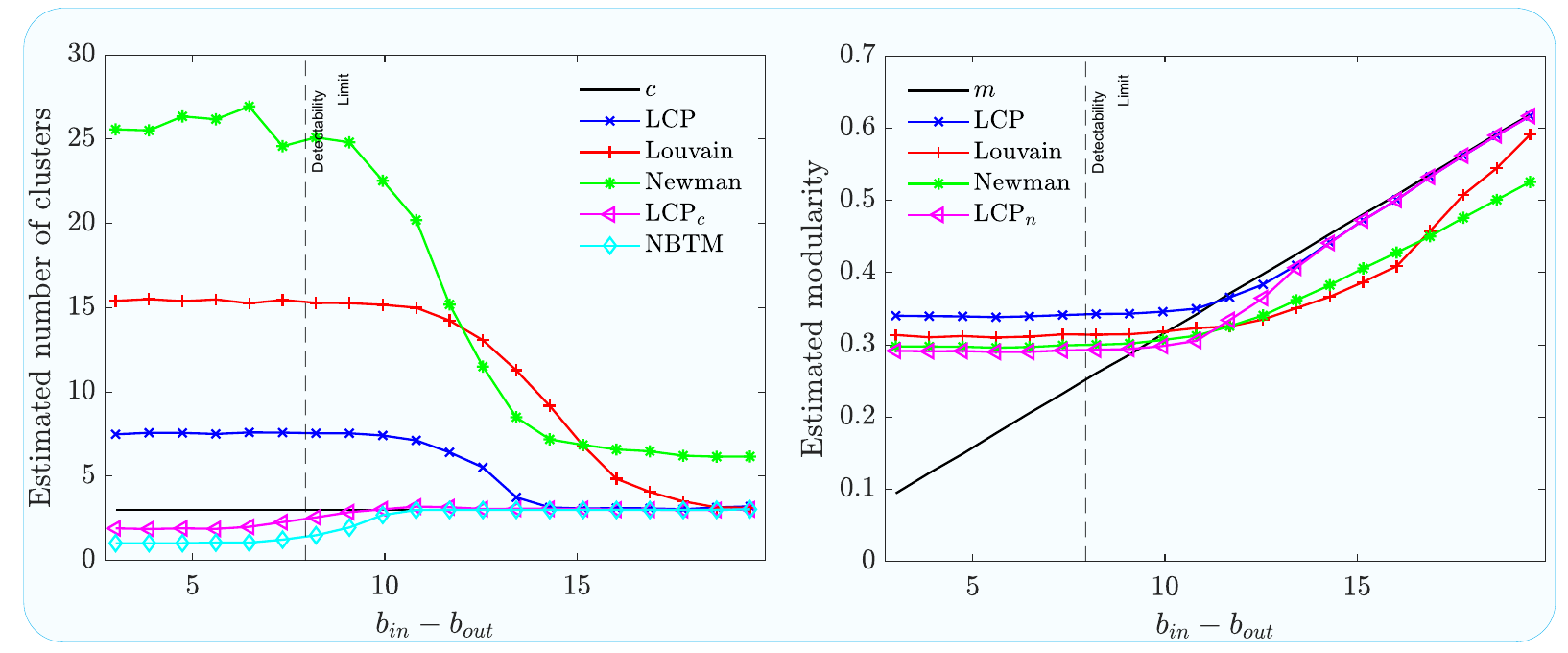}
		\caption{ The estimated number of clusters (left-figure) in SSBM graphs with $N=999$ nodes, average degree $d_{av}=7$ and $c=3$ clusters, for different values of parameters $b_{in}$ and $b_{out}$. The modularity of the estimated partitions is presented in right figure. The vertical dashed line indicates the clustering detectability threshold.}
		\label{Fig_Results_3_clusters}
	\end{center}
\end{figure*}

\begin{figure*}[!]
	\begin{center}
		\includegraphics[angle=0, scale= 1.00]{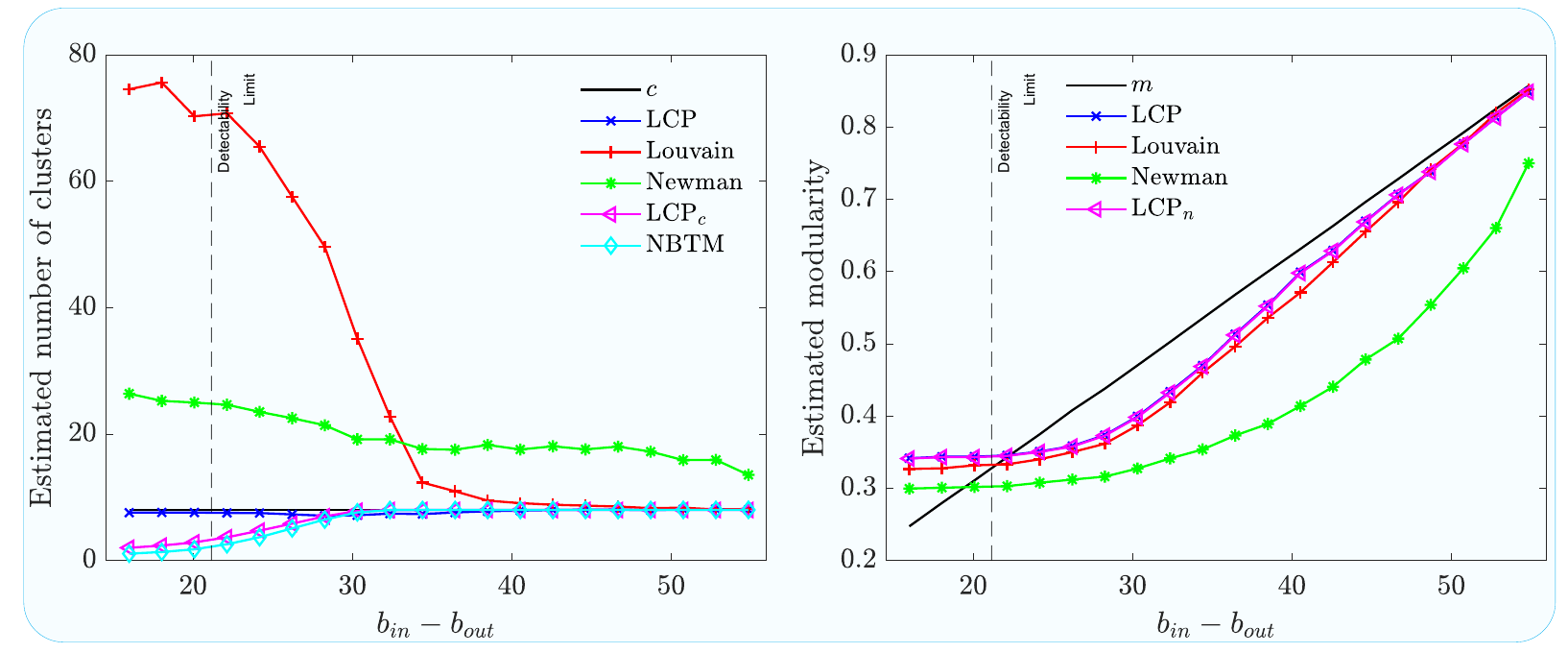}
		\caption{ The estimated number of clusters (left-figure) in SSBM graphs with $N=1000$ nodes, average degree $d_{av}=7$ and $c=8$ clusters, for different values of parameters $b_{in}$ and $b_{out}$. The modularity of the estimated partitions is presented in right figure. The vertical dashed line indicates the clustering detectability threshold.}
		\label{Fig_Results_8_clusters}
	\end{center}
\end{figure*}

We compare the clustering performance of our LCP with that of clustering methods introduced in Appendix  \ref{App_Clustering_Algorithms}, on a same graph generated by the symmetric stochastic block model (SSBM) with clusters of equal size. All graphs have $N=1000$ nodes. We vary the parameters $b_{in}$ and $b_{out}$ using (\ref{Eq:Expected_Degree}) in a way to keep the average degree $d_{av}=7$ fixed. For each SSBM network, we execute the clustering methods $10^{2}$ times and present the mean number of estimated clusters and mean modularity of produced partitions in Figures (\ref{Fig_Results_2_clusters}-\ref{Fig_Results_8_clusters}). While the non-back tracking algorithm (Sec \ref{Sec_non_back_tracking_matrix}) and our LCP$_{c}$ (Sec \ref{Non_back_tracking_as_a_process}) estimate only number of clusters, Newman's method (Sec \ref{Sec_Newman_modularity}), the Louvain method (Sec \ref{Sec_Louvain}) and our LCP (Sec \ref{Recursive_algorithm_sec}) and LCP$_{n}$ (Sec \ref{Recursive_algorithm_given_N_sec}) estimate both number of clusters and the cluster membership of each node. The attractive strength $\alpha=0.95$ and the repulsive strength $\delta = 10^{-3}$ are used in all simulations. Weights of $60\%$ links in total are scaled using (\ref{Ranking_Distance_eq}), evenly over $30$ iterations, where in $i$-th iteration scaled weight is $\frac{0.05\cdot i}{30}$.

The clustering performance on an SSBM graph with $c=2$ clusters  is presented in Figure \ref{Fig_Results_2_clusters}. The non-back tracking algorithm and our LCP$_{h}$ achieve the best performance in estimating the number of communities $c$, as shown in left-part of Figure \ref{Fig_Results_2_clusters}. Further, our LCP outperforms both modularity-based methods, in identifying the number of communities as well as in modularity $m$.

Figure \ref{Fig_Results_2_clusters} illustrates a significant difference in performance between our LCP and the non-back tracking matrix method. The modularity $m$ of our LCP is plotted in blue, while LCP$_c$, when the number $c$ of communities is known, is presented in magenta. Our LCP and the other two modularity-based methods perform poorly in recognising number $c$ of clusters for a wide range of $b_{in}-b_{out}$, because modularity-based methods generate partitions of higher modularity than that of the original network (in black colour), but with different number of communities!  

Our LCP$_{n}$ shows that even for a given number of clusters $c$, there exists a graph partition with a higher modularity than the modularity of the original network, within theoretically detectable regime.

In a graph with $c=3$ communities, Figure \ref{Fig_Results_3_clusters} shows that the non-back tracking matrix method again outperforms other three methods in identifying number of communities. Our LCP produces better partitions over the entire range $b_{in}-b_{out}$ than Newman's and the Louvain method. As in the case with $c=2$ clusters, by optimising modularity, the exact number of communities in SSBM graph cannot always be recovered, because there exist other partitions with higher number $c$ of clusters with a higher modularity.

Finally, Figure \ref{Fig_Results_8_clusters} shows a similar view for an SSBM graph with $c=8$ clusters. Our LCP again outperforms the other three methods in estimating number of communities over the entire range of $b_{in}-b_{out}$ values. In addition, the modularity of our LCP is superior to the other two methods.

\section{Conclusion}\label{Sec:Conclusion}

In this paper, we propose a linear clustering process (LCP) on a network consisting of an attraction and repulsion process between neighbouring nodes, proportional to how similar or different their neighbours are. Based on nodal positions, we are able to estimate both the number $c$ and the nodal membership of communities. Our LCP outperforms modularity-based clustering algorithms, such as Newman's and the Louvain method, while being of the same computational complexity. The proposed LCP allows estimating the number $c$ of clusters as accurately as the non-back tracking matrix.

The linear clustering process LCP is described by a matrix $I + W - \text{diag}(W\cdot u)$, which can be regarded as an operator acting on the position of nodes, comparable to quantum mechanics (QM). In QM, an operator describes a dynamical action on a set of particles. Since quantum mechanical operators are linear, the dynamics are exactly computed via spectral decomposition. In a same vein, our operator $I + W - \text{diag}(W\cdot u)$ is linear and describes via attraction and repulsion a most likely ordering of the position of nodes that naturally leads to clusters, via spectral decomposition, in particular, via the eigenvector $y_2$ in Section \ref{model_time_dependence_subsection}.

\section*{Acknowledgements}
The authors are grateful to S. Fortunato for useful comments.
This research is part of NExTWORKx, a collaboration between TU Delft and KPN on future communication networks.

\appendix
\section{Clustering algorithms}\label{App_Clustering_Algorithms}

\subsection{Louvain Method}\label{Sec_Louvain}

The Louvain method is a simple, yet powerful heuristic clustering algorithm, proposed by Blondel \textit{et al.} \cite{blondel2008fast}. The method is based on an iterative, unsupervised, two-step procedure that optimizes modularity $m$.
Initially, a directed graph $G$ with an $N\times N$ weighted adjacency matrix $M$ is partitioned in $N$ clusters, where each node constitutes its own cluster or community. 

In the first stage, the algorithm examines how the graph modularity $m$ changes if a node $i$ would be assigned to a community of its neighbouring node $j\in \mathcal{N}_{i}$. The modularity gain $\Delta m$ in case node $i$ is assigned to community $h$ of adjacent node $j$ has been determined in \cite{blondel2008fast} as
\begin{equation}\label{Eq:Louvain_Def}
	\begin{aligned}
		\Delta m = \left( \frac{\sum_{\text{in}} + 2\sum_{l:C_{lj}=1}M_{il}}{2L} - \left(\frac{\sum_{\text{tot}} + d_{i}}{2L}\right)^{2} \right) - \left( \frac{\sum_{\text{in}}}{2L} - \left(\frac{\sum_{\text{tot}}}{2L}\right)^{2} -  \left(\frac{d_{i}}{2L}\right)^{2} \right),
	\end{aligned}
\end{equation}
where the sum of the weights of intra-community links in $h$ is $\sum_\text{in}$, while $\sum_\text{tot}$ denotes the sum of the weights of all links in $G$ incident to any node in community $h$. Node $i$ is assigned to the community with the largest positive gain in modularity $m$. In case all computed gains $\Delta m$ are either negative or smaller than a predefined small positive threshold value, node $i$ remains in its original community. The first stage ends when modularity $m$ cannot be further increased by re-assigning nodes to communities of neighbours.

In the second stage of an iteration, the weighted graph from the first stage is transformed into a new weighted graph, where each community is presented by a node. The link weight between two nodes $h$ and $g$ equals the sum of weights of all links between communities $h$ and $g$ in the graph from the first stage. Furthermore, the weight of a self-loop of node $g$ in the new graph equals the sum of weights of all intra-community links in cluster $g$ of the graph from the previous stage. The new graph is provided to the first stage in the next iteration. The algorithm stops when modularity $m$ cannot be increased further. The time complexity of the Louvain method is linear in the number of links $O(L)$ on sparse graphs \cite{blondel2008fast}.

\subsection{Newman's Method of Optimal Modularity}\label{Sec_Newman_modularity}

Newman \cite{newman2006modularity} proposed a clustering algorithm that is based on modularity optimisation. The algorithm starts with estimating the bisection of a graph $G$, generating the highest modularity $m$ from (\ref{Eq_modularity}), that can be rewritten as follows:
\begin{equation}\label{Eq:Newman_modularity}
	m = \frac{1}{4L}y^{T}\cdot M \cdot y,
\end{equation}
where the $N\times 1$ vector $y$ is composed of values $1$ and $-1$, denoting cluster membership of each node, while the $N\times N$ modularity matrix $M = A - \frac{1}{2L}\cdot d\cdot d^{T}$ has the following eigenvalue decomposition
\begin{equation}\label{Eq:Newman_Eig_Dec}
	M = \sum\limits_{i=1}^{N} \zeta_{i}\cdot z_{i}\cdot z_{i}^{T},
\end{equation}
where the $N\times 1$ eigenvector $z_{i}$ corresponds to the $i$-th eigenvalue $\zeta_{i}$. Further, the vector $y = \sum_{j=1}^{N}(z_{j}^{T}\cdot y)\cdot z_{j}$ can be written as a linear combination of eigenvectors $\{z_{i}\}_{1\leq i \leq N}$, which transforms (\ref{Eq:Newman_modularity}) to
\begin{equation}\label{Eq:Newman_Mod_2}
	m = \frac{1}{4L}\sum\limits_{i=1}^{N}\zeta_{i}\cdot (z_{j}^{T}\cdot y)^{2}.
\end{equation}
In order to maximise the modularity $m$, Newman \cite{newman2006modularity} proposed to define $y_{i}=1$, in case $(z_{1})_{i}>0$, otherwise $y_{i} = -1$. In a next iteration, the same procedure of spectral division into two partitions is repeated on both sub-graphs. However, using only the block sub-matrix of $M$, corresponding to cluster $g$ in next iteration would not take into account inter-community links. Instead, for the estimated cluster $g$, the modularity matrix $M_{g}$ is updated as
\begin{equation}\label{Eq:Newman_Updating_M}
	M_{g} = m_{ij} - \left(\sum_{k\in g}m_{ik}\right)\cdot \delta_{ij},
\end{equation}
where Kronecker delta $\delta_{ij}=1$ if $i=j$, otherwise $\delta_{ij}=0$. The algorithm stops when the modularity $m$ cannot be further improved.

\subsection{Non-back tracking matrix}\label{Sec_non_back_tracking_matrix}

The non-back tracking clustering method estimates the number of clusters in a network, based on the spectrum of the non-back tracking matrix $B$, that contains information about $2$-hop directed walks in a network $G$, that are not closed \cite{krzakala2013spectral}. Given an undirected network $G(\mathcal{N},\mathcal{L})$, for each link $i \sim j$ between nodes $i$ and $j$, two directed links $(i \rightarrow j)$ and $(j \rightarrow i)$ are created. By transforming each link in $G$ into a bi-directional link pair, we compose in total $2L$ links. The $2L\times 2L$ non-back tracking matrix $B$ is defined as follows:
\begin{equation}\label{Eq:Non-back tracking B}
	B_{(u \rightarrow v),(w \rightarrow z)} = \begin{cases}
		1 & \text{if $v = w$ and $u\ne z$}\\
		0 & \text{otherwise},
	\end{cases}
\end{equation}
where $v,w,z \in \mathcal{N}$. Since the non-back tracking matrix $B$ is asymmetric, its eigenvalues are generally complex. Furthermore, a vast majority of eigenvalues lie within a circle in complex plain, with centre at the origin and with radius equal to the square root of the largest eigenvalue. Krzakala \textit{et al.} \cite{krzakala2013spectral} hypothesized that the number of clusters in $G$ equals the number of real-valued eigenvalues outside the circle. Computing the eigenvalues of the non-back tracking matrix $B$ is of computational complexity $O(L^{3})$. However, the complexity can be reduced to $O(N^{3})$, as explained in \cite[p. 20]{budel2020detecting}. The non-back tracking matrix method is denoted as NBTM in Section  \ref{Sec_Results}.

\section{LCP in continuous time}

\label{Sec_Continuous_time} We explain the physical intuition of our
clustering process in continuous time $t$, where the position $x_{i}(t)$ of a
node $i$ is assumed to change continuously with time $t$. The change
$x_{i}(t+\Delta t)-x_{i}(t)$ in position of node $i$ at time $t$ for small
increments $\Delta t$ is proportional to the sum over neighbours $j$ of the
difference $x_{j}(t)-x_{i}(t)$ in position weighted by the resultant force
between attraction and repulsion:%
\begin{equation}\label{law_continuous_time}
\frac{dx_{i}(t)}{dt}=\sum\limits_{j\in\mathcal{N}_{i}}\bigg(\frac{\alpha\cdot\left(\left\vert \mathcal{N}_{j}\cap\mathcal{N}_{i}\right\vert +1\right)}{d_{j}d_{i}} -\frac{\frac{1}{2} \cdot\delta\cdot\left(  \left\vert \mathcal{N}_{j}\setminus \mathcal{N}_{i}\right\vert +\left\vert \mathcal{N}_{i}\setminus\mathcal{N}_{j}\right\vert -2\right)}{d_{j}d_{i}}\bigg)\cdot \Big(x_{j}(t)-x_{i}(t)\Big)
\end{equation}
where $\alpha$ and $\delta$ are, in the continuous-time setting, the rates
(with units s$^{-1}$) for attraction and repulsion, respectively. The law
(\ref{law_continuous_time}) of the nodal positioning $x_{i}\left(  t\right)  $
for each $i\in\mathcal{N}$ deviates from physical repulsion between charged
particles, where the force is proportional to $\left(  x_{j}(t)-x_{i}%
(t)\right)  ^{-b}$ for some positive number $b$. The important advantage of
the law (\ref{law_continuous_time}) is its linearity that allows an exact
mathematical treatment. The linear dynamic process (\ref{law_continuous_time}) is proportional to the underlying graph, which we aim to cluster; a non-linear law depends intricately on the underlying graph and may result in a lesser clustering. The drawback of the linear dynamic process (\ref{law_continuous_time}), as investigated below in Section
\ref{model_time_dependence_subsection}, lies in the steady state, where the
attractive and repulsive forces are precisely in balance.   

After dividing both sides by $\delta$,
\[\frac{dx_{i}(t)}{d\left(\delta t\right)} = \sum\limits_{j\in\mathcal{N}_{i}}\bigg(\frac{\frac{\alpha}{\delta}\cdot\left(  \left\vert \mathcal{N}_{j} \cap\mathcal{N}_{i}\right\vert +1\right)}{d_{i}d_{j}} - \frac{\frac{1}{2}\cdot\left(  \left\vert \mathcal{N}_{j}\setminus\mathcal{N}_{i}\right\vert +\left\vert \mathcal{N}_{i}\setminus\mathcal{N}_{j}\right\vert -2\right)}{d_{i}d_{j}}\bigg) \cdot\Big(x_{j}(t)-x_{i}(t)\Big)
\]
and defining the normalized time by $t^{\ast}=\delta t$ and the effective attraction rate $\tau=\frac{\alpha}{\delta}$, the governing equation (\ref{law_continuous_time}) reduces to
\begin{equation}\label{law_continuous_time_normalized}
\frac{dx_{i}(t^{\ast})}{dt^{\ast}} =  \sum\limits_{j\in\mathcal{N}_{i}} \Bigg(\frac{\tau\cdot\left(  \left\vert \mathcal{N}_{j}\cap\mathcal{N}_{i}\right\vert +1\right)}{d_{i}d_{j}} - \frac{\left(  \frac{\left\vert \mathcal{N}_{j}\setminus\mathcal{N}_{i}\right\vert + \left\vert \mathcal{N}_{i}\setminus\mathcal{N}_{j}\right\vert}{2}-1\right)}{d_{i}d_{j}} \Bigg) \cdot \Big(x_{j}(t^{\ast})-x_{i}(t^{\ast})\Big)
\end{equation}
The position $x_{i}(t^{\ast})$ of node $i$ is now expressed in the
dimensionless time $t^{\ast}$, where the actual time $t=\frac{t^{\ast}}%
{\delta}$ is measured in units of $\frac{1}{\delta}$. By scaling or
normalizing the time, the repulsion strength or rate $\delta$ has been
eliminated, illustrating that the clustering process only depends upon one
parameter, the effective attraction rate $\tau$. Relation
(\ref{eq_degree_in_set_notation}) indicates that the weight of the position
difference
\[
w_{ij}=\frac{\tau\cdot\left(  \left\vert \mathcal{N}_{j}\cap\mathcal{N}%
	_{i}\right\vert +1\right)  -\left(  \frac{\left\vert \mathcal{N}_{j}%
		\setminus\mathcal{N}_{i}\right\vert +\left\vert \mathcal{N}_{i}\setminus
		\mathcal{N}_{j}\right\vert }{2}-1\right)  }{d_{i}d_{j}}%
\]
lies in the interval $\left(  -\frac{\frac{d_{i}+d_{j}}{2}-1}{d_{i}d_{j}}%
,\frac{\tau}{d_{i}}\right)  $ and that the elements
$w_{ij}=w_{ji}$ define the symmetric $N\times N$ weight matrix $W$, which is
specified in (\ref{def_martrix_W}) below. Although symmetry is physically not
required
\footnote{The process described by
	\[
	\frac{dx_{i}(t)}{dt}=\sum\limits_{j\in\mathcal{N}_{i}}\frac{\alpha\cdot\left(
		\left\vert \mathcal{N}_{j}\cap\mathcal{N}_{i}\right\vert +1\right)
		-\delta\cdot\left(  \left\vert \mathcal{N}_{j}\setminus\mathcal{N}%
		_{i}\right\vert -1\right)  }{d_{j}d_{i}}\cdot\Big(x_{j}(t)-x_{i}(t)\Big)
	\]
	also works.},
the analysis below is greatly simplified, because eigenvalues
and eigenvectors of a symmetric matrix are real.

We rewrite the law (\ref{law_continuous_time_normalized}) as
\begin{eqnarray*}
	\frac{dx_{i}(t^{\ast})}{dt^{\ast}}=\sum\limits_{j\in\mathcal{N}_{i}} w_{ij}x_{j}(t^{\ast})-x_{i}(t^{\ast})\sum\limits_{j\in\mathcal{N}_{i}} w_{ij} = \sum_{j=1}^{N}a_{ij}w_{ij}x_{j}(t^{\ast})-x_{i}(t^{\ast})v_{i}
\end{eqnarray*}
where $v_{i}=\sum\limits_{j\in\mathcal{N}_{i}}w_{j}=\sum_{j=1}^{N}a_{ij}%
w_{ij}$ is independent of time $t^{\ast}$. The cluster positioning law
(\ref{law_continuous_time}) for the vector $x(t^{\ast})$ in continuous time
is, in matrix form,
\begin{equation}
	\frac{dx(t^{\ast})}{dt^{\ast}}=(A\circ W-\text{diag}(v))x(t^{\ast
	})\label{law_continuous_time_matrix}%
\end{equation}
where the Hadamard product \cite{horn2012matrix} is denoted by $\circ$ and the vector $v=(A\circ W)u$. The
corresponding solution of (\ref{law_continuous_time_matrix}) is \cite[eq.
(6)]{van2016approximate}
\begin{equation}
	x(t^{\ast})=e^{(A\circ W-\text{diag}((A\circ W)u))t^{\ast}}%
	x(0)\label{solution_position_vector_continuous_time}%
\end{equation}
which illustrates that a steady state is reached, provided that the real part of the largest eigenvalue of the matrix $H=(A\circ W-$diag$((A\circ W)u))$ is not positive. 

\section{Proof of Theorems}
\subsection{Proof of Theorem 
	\ref{clustering_law_additive_matrix_form}}\label{App_A}

Similarly as in Section \ref{Sec_Continuous_time}, we rewrite the sum over all neighbours in the governing equation (\ref{clustering_law_discrete_time}) in terms of the elements of the $N\times N$ adjacency matrix $A$:
\begin{equation}\label{gov_eq_node_level_extended}
x_{i}[k+1] - x_{i}[k] = \sum\limits_{j=1}^{N} \frac{a_{ij}}{d_{i} d_{j}} \Big(x_{j}[k] - x_{i}[k]\Big) \Big( \alpha \left|\mathcal{N}_{j} \cap \mathcal{N}_{i} \right| - \frac{1}{2} \delta \left(\left|\mathcal{N}_{j} \setminus \mathcal{N}_{i} \right| + \left|\mathcal{N}_{i} \setminus \mathcal{N}_{j} \right|\right) \Big).
\end{equation}
Firstly, we denote the $N\times1$ vector $\tilde{d}=\Delta^{-1}\cdot u$ composed of the inverse nodal degrees:
\begin{equation}
	\tilde{d}=%
	\begin{bmatrix}
		\frac{1}{d_{1}} & \frac{1}{d_{2}} & \dots & \frac{1}{d_{N}}%
	\end{bmatrix}
	^{T} \label{Eq_deg_inverse_vector}%
\end{equation}
In the sequel, we will deduce the corresponding matrix form of
(\ref{gov_eq_node_level_extended}). With (\ref{eq_degree_in_set_notation}) and
(\ref{common_neig_set_not}), the degree $d_{i}$ of node $i$ distracted by the
number of common neighbours between nodes $i$ and $j$
(\ref{common_neig_set_not}), equals the number of node $i$ neighbours, not
adjacent to node $j$:
\begin{equation}
	\label{diff_neig_set_not_i}\left|  \mathcal{N}_{i} \setminus\mathcal{N}_{j}
	\right|  = \left(  d\cdot u^{T} - A^{2}\right)  _{ij}.
\end{equation}
Similarly, the number of node $j$ neighbours that do not share link with node
$i$ has following matrix form:
\begin{equation}
	\label{diff_neig_set_not_j}\left|  \mathcal{N}_{j} \setminus\mathcal{N}_{i}
	\right|  = \left(  u\cdot d ^{T} - A^{2}\right)  _{ij}.
\end{equation}
Finally, the position difference $(x_{j}[k]-x_{i}[k])$ between nodes $i$ and
$j$ at time $k$ equals the $ij$-th element of the matrix below:
\begin{equation}
	\label{position_diff_matrix_notation}\left(  x_{j}[k] - x_{i}[k]\right)  =
	\left(  u\cdot x^{T}[k] - x[k]\cdot u^{T}\right)  _{ij},
\end{equation}
while dividing by node $i$ ($j$) degree $d_{i}$ ($d_{j}$) is equivalent to
product with the $ij$-th element of the $N\times N$ matrix $(\tilde{d}\cdot u^{T})_{ij}$ and $(u\cdot\tilde{d}^{T})_{ij}$, respectively. By implementing
matrix notations (\ref{common_neig_set_not}), (\ref{diff_neig_set_not_i}),
(\ref{diff_neig_set_not_j}) and (\ref{position_diff_matrix_notation}) into the governing equation (\ref{gov_eq_node_level_extended}) and by applying the distributive property of the Hadamard product  \cite[p. 477]{horn2012matrix} we obtain:
\begin{equation}\label{gov_eq_matrix_not_3}
\begin{aligned} x[k+1] -x[k] = 
& \bigg(\left(u\cdot x^{T}[k] - x[k]\cdot u^{T}\right) \circ A\circ  \\
& \left(u \cdot \tilde{d}^{T}\right) \circ \left(\tilde{d} \cdot u^{T}\right) \circ \Big(\left(\alpha + \delta\right) \cdot \left(A^{2} + A\right) - \\ 
& \frac{1}{2} \delta \cdot \left(u\cdot d^{T} + d\cdot u^{T }\right)\Big)\bigg)\cdot u. 
\end{aligned}
\end{equation}
We define the $N\times N$ topology-based matrix $W$ as follows:
\begin{equation}\label{eq_matrix_W_def}
W = A\circ\left(  u \cdot\tilde{d}^{T}\right)
\circ\left(  \tilde{d} \cdot u^{T}\right)  \circ \Big(\left(  \alpha+ \delta\right)  \cdot\left(  A^{2} + A\right)  - \frac{1}{2}\delta\left( u\cdot d^{T} + d\cdot u^{T }\right)  \Big).
\end{equation}
Using the distributive property of a Hadamard product  \cite[p. 477]{horn2012matrix}, we develop the equation (\ref{eq_matrix_W_def}) further:
\begin{equation}\label{eq_matrix_W_s1}%
\begin{split}
W =  &  \left(  \alpha+ \delta\right)  \cdot\Big(\left(  u \cdot\tilde{d}%
^{T}\right)  \circ\left(  \tilde{d} \cdot u^{T}\right)  \circ A\circ\left( A^{2} + A\right)  \Big) -\\
&  \frac{1}{2}\delta\Big(A\circ\left(  u \cdot\tilde{d}^{T}\right)  \circ\left( \tilde{d} \cdot u^{T}\right)  \circ\left(  u\cdot d^{T}\right)  \Big) -\\
&  \frac{1}{2}\delta\Big(A\circ\left(  u \cdot\tilde{d}^{T}\right)  \circ\left( \tilde{d} \cdot u^{T}\right)  \circ\left(  d\cdot u^{T}\right)  \Big).
\end{split}
\end{equation}
Since the Hadamard product is commutative \cite[p. 477]{horn2012matrix}, we can reorder the products in previous equation. The Hadamard product
$\left(  u \cdot\tilde{d}^{T}\right)  \circ\left(  u\cdot d^{T}\right)$ equals all-one matrix $J$. Similarly, the product $\left( \tilde{d} \cdot u^{T}\right)
\circ\left(  d\cdot u^{T}\right)  = J$ . We further transform the Hadamard product of $\left(  A\circ A^{2}+A\right)  $ and the outer
products $\left(  u \cdot\tilde{d}^{T}\right)  $ and $\left(  \tilde{d} \cdot u^{T}\right)  $ into product with diagonal matrices $\Delta^{-1}\cdot\left(  A\circ A^{2}+A\right)  \cdot \Delta^{-1}$. Thus, equation (\ref{eq_matrix_W_s1}) transforms to (\ref{def_martrix_W}).
Substituting (\ref{eq_matrix_W_def}) into (\ref{gov_eq_matrix_not_3}) yields
\begin{equation}
	\label{gov_eq_matrix_not_4}x[k+1] - x[k] = \bigg(\left(  u\cdot x^{T}[k] -
	x[k]\cdot u^{T}\right)  \circ W\bigg)\cdot u.
\end{equation}
The Hadamard product of a matrix with an outer product of two vectors is equivalent to the product with diagonal matrices of vectors composing the outer product \cite[p. 477]{horn2012matrix}. 
Thus, we further transform the governing equation
(\ref{gov_eq_matrix_not_4}):
\begin{equation}
	x[k+1] - x[k] = W\cdot\text{diag}\left(
	x[k]\right)  \cdot u - \text{diag}\left(  x[k]\right)  \cdot\left(  W\cdot
	u\right)  ,
\end{equation}
where the last term $\text{diag}\left(  x[k]\right)  \cdot\left(  W\cdot
u\right)  $ represents the Hadamard product of two vectors, $x[k] \circ\left(
W\cdot u\right)  $ and can be presented as $\text{diag}\left(  W\cdot
u\right)  \cdot x[k]$. Thus, the equation transforms into (\ref{clustering_law_matrix_equation})
which completes the proof. $\hfill\square$

\subsection{ Proof of Property \ref{prop_M_eigenvalues}}
\label{sec_prop_M_eigenvalues}
We observe that
\[
W\cdot u-\text{diag}\left(  W\cdot u\right)  \cdot u=0
\]
implying that the all-one vector $u$ is an eigenvector of the matrix
$W-\text{diag}\left(  W\cdot u\right)  $ belonging to the zero eigenvalue.
Therefore, the $N\times N$ matrix $I+W-\text{diag}\left(  W\cdot u\right)  $
has an eigenvalue $1$ corresponding to the all-one vector $u$.

By the Perron-Frobenius theorem \cite{van2010spectral} for a non-negative matrix, the principal
eigenvector, belonging to the largest eigenvalue, has non-negative components. Since the eigenvector $u$ has non-negative components and all eigenvectors of
a symmetric matrix are orthogonal, it follows that the all-one vector $u$ is
the Perron or principal eigenvector belonging to the largest eigenvalue 1 of the matrix
$I+W-\text{diag}\left(  W\cdot u\right)  $ and, thus, all other real
eigenvalues are, in absolute value, smaller than 1.  $\hfill\square$

\subsection{Proof of Property
	\ref{prop_alpha_delta}}\label{sec_prop_alpha_delta}

The non-negativity of the matrix $I+W-\text{diag}\left(  W\cdot u\right)
$ implies that $w_{ij}\geq0$ for $i\neq j$ and $1+w_{ii}-\sum_{k=1}^{N}%
w_{ik}\geq0$, hence,
\[
1\geq\sum_{k=1;k\neq i}^{N}w_{ik}\geq0
\]
Equivalently, the symmetric matrix $W-\text{diag}\left(  W\cdot u\right)  $
has positive off-diagonal elements, but negative diagonal elements, similarly
to the infinitesimal generator of a Markov chain (which is minus a weighted
Laplacian \cite{van2014performance}). Introducing the explicit expression
(\ref{def_matrix_W_element}) and requiring that each element $w_{ij}$ is
non-negative,
\[w_{ij}=a_{ij}\frac{\alpha\cdot\left(  \left\vert \mathcal{N}_{j}%
		\cap\mathcal{N}_{i}\right\vert +1\right)  -\delta\cdot\left(  \frac{\left\vert
			\mathcal{N}_{j}\setminus\mathcal{N}_{i}\right\vert +\left\vert \mathcal{N}%
			_{i}\setminus\mathcal{N}_{j}\right\vert }{2}-1\right)  }{d_{i}d_{j}}\geq0
\]
leads to
\[
\frac{\alpha}{\delta}\geq\frac{1}{2}\cdot\frac{\left(  \left\vert
	\mathcal{N}_{j}\setminus\mathcal{N}_{i}\right\vert +\left\vert \mathcal{N}%
	_{i}\setminus\mathcal{N}_{j}\right\vert -2\right)  }{\left(  \left\vert
	\mathcal{N}_{j}\cap\mathcal{N}_{i}\right\vert +1\right)  }%
\]
which holds for any $i,j\neq i\in\mathcal{N}$. With
(\ref{eq_degree_in_set_notation}), (\ref{common_neig_set_not}) and
$d_{i}=(Au)_{i}=(A^{2})_{ii}$, the condition for the ratio $\frac{\alpha
}{\delta}$ becomes
\[\frac{\alpha}{\delta}\geq\max_{i,j\neq i\in\mathcal{N}}\frac{1}{2}\cdot
	\frac{\left(  \left\vert \mathcal{N}_{j}\setminus\mathcal{N}_{i}\right\vert +\left\vert \mathcal{N}_{i}\setminus\mathcal{N}_{j}\right\vert -2\right) }{\left(  \left\vert \mathcal{N}_{j}\cap\mathcal{N}_{i}\right\vert +1\right)}=\max_{i,j\neq i\in\mathcal{N}}\frac{d_{i}+d_{j}}{2\left( (A^{2})_{ij}+1\right)}-1
\]
which simplifies to
\begin{equation}
	\frac{\alpha}{\delta}\geq d_{\max}-1\label{condition_1}%
\end{equation}

We write $\sum_{k=1;k\neq i}^{N}w_{ik}$ with (\ref{def_matrix_W_element}) as
\[\sum_{k=1;k\neq i}^{N}w_{ik}= \frac{1}{d_{i}}\sum_{k=1}^{N}a_{ik}\bigg(\frac
	{\alpha\cdot\left(  \left\vert \mathcal{N}_{k}\cap\mathcal{N}_{i}\right\vert
		+1\right)}{d_{k}} -\frac{\frac{\delta}{2}\cdot\left(  \left\vert \mathcal{N}_{k}%
		\setminus\mathcal{N}_{i}\right\vert +\left\vert \mathcal{N}_{i}\setminus
		\mathcal{N}_{k}\right\vert -2\right)  }{d_{k}}\bigg)
\]
Introducing (\ref{eq_degree_in_set_notation}) and (\ref{common_neig_set_not}),
\[\begin{aligned}
		\sum_{k=1;k\neq i}^{N}w_{ik} &  =\frac{1}{d_{i}}\sum_{k=1}^{N}a_{ik}%
		\frac{\alpha\cdot\left(  (A^{2})_{ik}+1\right)  -\frac{\delta}{2}\cdot\left(
			d_{i}+d_{k}-2\left(  (A^{2})_{ik}+1\right)  \right)  }{d_{k}}\\
		&  =\frac{\alpha}{d_{i}}\sum_{k=1}^{N}a_{ik}\frac{\left(  (A^{2}%
			)_{ik}+1\right)  }{d_{k}}-\frac{\delta}{2d_{i}}\sum_{k=1}^{N}a_{ik}\left(
		\frac{d_{i}}{d_{k}}+1-\frac{2\left(  (A^{2})_{ik}+1\right) }{d_{k}}\right)\end{aligned}\]
leads, with $d_{i}=\sum_{k=1}^{N}a_{ik}$, to
\[\begin{aligned}
		\sum_{k=1;k\neq i}^{N}w_{ik}=\frac{\alpha+\delta}{d_{i}}\sum_{k=1}^{N}%
		a_{ik}\frac{\left(  (A^{2})_{ik}+1\right)  }{d_{k}}-\frac{\delta}{2}%
		-\frac{\delta}{2}\sum_{k=1}^{N}\frac{a_{ik}}{d_{k}}\end{aligned}
\]
The second condition $\sum_{k=1;k\neq i}^{N}w_{ik}\leq1$,%
\[
\frac{\alpha+\delta}{d_{i}}\sum_{k=1}^{N}a_{ik}\frac{\left(  (A^{2}%
	)_{ik}+1\right)  }{d_{k}}-\frac{\delta}{2}-\frac{\delta}{2}\sum_{k=1}^{N}%
\frac{a_{ik}}{d_{k}}\leq1
\]
must hold for all $i\in\mathcal{N}$, which translates to%
\[\begin{aligned}
		1  & \geq\max_{i\in\mathcal{N}}\left(  \frac{\alpha+\delta}{d_{i}}\sum
		_{k=1}^{N}a_{ik}\frac{\left(  (A^{2})_{ik}+1\right)  }{d_{k}}-\frac{\delta}%
		{2}-\frac{\delta}{2}\sum_{k=1}^{N}\frac{a_{ik}}{d_{k}}\right)  \\
		& \geq\left(  \alpha+\delta\right)  \max_{i\in\mathcal{N}}\frac{1}{d_{i}}%
		\sum_{k=1}^{N}a_{ik}\frac{\left(  (A^{2})_{ik}+1\right)  }{d_{k}}-\frac
		{\delta}{2}-\frac{\delta}{2}\min_{i\in\mathcal{N}}\sum_{k=1}^{N}\frac{a_{ik}%
		}{d_{k}}%
	\end{aligned}
\]

With $\left(  (A^{2})_{ik}+1\right)  \leq d_{k}$ if $a_{ik}=1$, we have
$\frac{1}{d_{i}}\sum_{k=1}^{N}a_{ik}\frac{\left(  (A^{2})_{ik}+1\right)
}{d_{k}}\leq1$, while $\frac{d_{i}}{d_{\min}}\geq\sum_{k=1}^{N}\frac{a_{ik}%
}{d_{k}}\geq\frac{d_{i}}{d_{\max}}$. Hence, the second condition becomes%
\begin{equation}
	1\geq\alpha+\frac{\delta}{2}\left(  1-\frac{d_{\min}}{d_{\max}}\right)
	\label{condition_2}%
\end{equation}
illustrating that $\alpha\leq1$. Combining the two conditions
(\ref{condition_1}) and (\ref{condition_2}) into a linear set of inequalities%
\[
\left\{
\begin{array}
	[c]{c}%
	0\geq-\alpha+\delta\left(  d_{\max}-1\right)  \\
	1\geq\alpha+\frac{\delta}{2}\left(  1-\frac{d_{\min}}{d_{\max}}\right)
\end{array}
\right.
\]
or in matrix form, where $\curlyeqsucc$ denotes componentwise inequalities
\cite[p. 32, 40]{boyd2004convex}
\[
\left[
\begin{array}
	[c]{c}%
	0\\
	1
\end{array}
\right]  \curlyeqsucc\left[
\begin{array}
	[c]{cc}%
	-1 & \left(  d_{\max}-1\right)  \\
	1 & \frac{1}{2}\left(  1-\frac{d_{\min}}{d_{\max}}\right)
\end{array}
\right]  \left[
\begin{array}
	[c]{c}%
	\alpha\\
	\delta
\end{array}
\right]
\]
yields, after inversion, the bounds (\ref{bounds_alpha}) and (\ref{bounds_delta}). $\hfill\square$

\bibliography{20220923_Linear_Clustering_Process_ArXiv}{}
\bibliographystyle{ieeetr}   

\end{document}